\documentclass[12pt]{article}
\usepackage[utf8]{inputenc}
\usepackage{color}
\usepackage{authblk}
\usepackage{latexsym}
\usepackage{epsfig,amssymb,euscript, mathrsfs,cite}
\usepackage{amsmath}
\usepackage{bbold}
\usepackage{mathrsfs} 
\usepackage{verbatim}
\usepackage{enumerate}
\definecolor{MyDarkBlue}{rgb}{0.15,0.15,0.45}
\usepackage[linktocpage=true]{hyperref}
\hypersetup{
colorlinks=true,
citecolor=MyDarkBlue,
linkcolor=MyDarkBlue,
urlcolor=MyDarkBlue,
pdfauthor={},
pdftitle={},
pdfsubject={hep-th}
}
\newsavebox{\ns}
\newsavebox{\dbrane}
\newsavebox{\dbshort}

\def\be{\begin{equation}}
\def\ee{\end{equation}}
\def\bea{\begin{eqnarray}}
\def\eea{\end{eqnarray}}

\newcommand{\nn}{\nonumber}

\newcommand\cF{\mathcal{F}}
\newcommand\cA{\mathcal{A}}
\newcommand{\ma}{\mathrm{a}}

\newcommand\R{\mathbb{R}}
\newcommand\Z{\mathbb{Z}}

\newcommand\C{\mathbb{C}}

\newcommand\diff{\mathrm{d}}

\newcommand{\dd}{\mathrm{d}}
\newcommand{\ii}{\mathrm{i}}

\newcommand{\ex}{\mathrm{e}}
\newcommand{\vol}{\mathrm{vol}}

\def\beq{\begin{equation}}
\def\eeq{\end{equation}}
\def\bea{\begin{align}} 
\def\eea{\end{align}}

\newlength{\sswidth}
\newcommand{\sla}[1]{
   \settowidth{\sswidth}{$#1$}
   \mbox{$\not{\hspace*{-0.15\sswidth}#1}$}}

\newcommand{\hook}{\mathbin{\rule[.2ex]{.4em}{.03em}\rule[.2ex]{.03em}{.9ex}}}

\newcommand{\Xb}{X_1}
\newcommand{\Xs}{X_2}

\newcommand{\gc}{\mathfrak{g}}

\newcommand{\met}{g}
\newcommand{\Met}{G}
\newcommand{\B}{Y}

\newcommand{\mexp}{\mathtt{g}}
\newcommand{\cutoff}{\delta}
\newcommand{\los}{\varepsilon}
\newcommand{\te}{\mathtt{e}}
\newcommand{\BJ}{\mathcal{J}}
\newcommand{\Jb}{\mathrm{J}}

\newcommand{\Jcur}{\mathscr{J}}
\newcommand{\mc}[1]{\mathcal{#1}}
\newcommand{\ph}[1]{\phantom{#1}}
\newcommand{\BK}{\mathcal{K}}
\newcommand{\Ex}{\mathrm{E}}

\newcommand{\aIfive}{a^I_2}

\newcommand{\mafive}{\ma_2}

\renewcommand{\j}{\varphi}
\newcommand{\Sren}{\mathbb{S}}
\newcommand{\CC}{+}
\newcommand{\HH}{-}
\newcommand{\bdrye}{{\rm e}}

\newcommand{\identity}{\mathbb{1}}

\newcommand{\st}{\varphi}

\numberwithin{equation}{section}       

\begin{document}

\begin{titlepage}

\begin{center}

\today

\vskip 2.0 cm 


{\Large \bf Gravitational free energy
 in topological AdS/CFT}

\vskip 15mm

{Pietro Benetti Genolini${}^a$, Paul Richmond${}^b$ and James Sparks${}^a$}

\vspace{1cm}
\centerline{${}^a${\it Mathematical Institute, University of Oxford,}}
\centerline{{\it Andrew Wiles Building, Radcliffe Observatory Quarter,}}
\centerline{{\it Woodstock Road, Oxford, OX2 6GG, UK}}
\vspace{1cm}
\centerline{${}^b${\it Dipartimento di Fisica, Universit\`a di Milano--Bicocca,}}
\centerline{{\it Piazza della Scienza 3, I-20126 Milano, Italy}}
\centerline{{and }}
\centerline{{\it INFN, sezione di Milano-Bicocca}}

\vskip .8cm 

{
\small \tt Pietro.BenettiGenolini@maths.ox.ac.uk, sparks@maths.ox.ac.uk, 
}\\
{\small \tt Paul.Richmond@mib.infn.it}

\end{center}

\bigskip
\begin{center}
{\bf {\sc Abstract}} 
\end{center}
We define and study a holographic dual to the topological twist of $\mathcal{N}=4$ gauge theories on Riemannian three-manifolds. 
The gravity duals are solutions to four-dimensional $\mathcal{N}=4$  gauged supergravity, where the three-manifold arises 
as a conformal boundary. Following our previous work, we show that the renormalized gravitational free energy of such solutions is independent of the boundary three-metric, as required for a topological theory. We then go further, analyzing the geometry of supersymmetric bulk solutions. 
Remarkably, we are able to show that the gravitational free energy of any smooth four-manifold filling of any three-manifold is 
always zero. Aided by this analysis, we prove a similar result for topological AdS$_5$/CFT$_4$. We comment on the implications 
of these results for the large $N$ limits of topologically twisted gauge theories in three and four dimensions, including
the ABJM theory and $\mathcal{N}=4$ $SU(N)$ super-Yang--Mills, respectively.
\end{titlepage}

\pagestyle{plain}
\setcounter{page}{1}
\newcounter{bean}
\baselineskip18pt
\tableofcontents

\allowdisplaybreaks

\section{Introduction and outline}\label{SecIntroduction}

The AdS/CFT correspondence conjectures an equivalence between certain quantum field theories (QFTs) and quantum gravity 
with appropriate boundary conditions \cite{Maldacena:1997re, Witten:1998qj, Gubser:1998bc}. 
In \cite{BenettiGenolini:2017zmu} we proposed to formulate a ``topological'' version of AdS/CFT, where the boundary 
theory is a topological QFT (TQFT).  In the dual gravity description this amounts to studying a more specific class of boundary conditions, which induce a Witten-type topological twist \cite{Witten:1988ze} of the dual QFT on the conformal boundary. Such TQFTs typically have a finite number of degrees of freedom, and in some instances can be solved completely.\footnote{For example, the Donaldson--Witten twist of $\mathcal{N}=4$ $SU(N)$ super-Yang--Mills is relevant for the set up in \cite{BenettiGenolini:2017zmu}.  For $N=2$ the topological correlation functions have been computed
explicitly for simply-connected spin four-manifolds of simple type in \cite{Labastida:1998sk}; they may be written in terms 
of Abelian Seiberg--Witten invariants.} Of course these theories are often also of independent mathematical interest, since observables 
are topological/diffeomorphism invariants.

A key motivation for studying AdS/CFT in this set up is that the field theory 
is potentially under complete control: observables are mathematically well-defined and exactly computable. One can then focus 
on the dual gravitational description. In principle this is defined by a quantum gravity path integral, with boundary conditions 
determined by the observable one is computing. However, we have no precise definition of this, and in practice an appropriate strong coupling (usually large rank $N$) limit of the QFT is described by supergravity. This classical limit is to be understood as a saddle point approximation to the quantum gravity path integral, 
where one instead finds classical solutions to supergravity with the appropriate boundary conditions. But
in general even this is quite poorly understood: which saddle point solutions should be included? For example, in addition to smooth real solutions, should one allow 
for certain types of singular and/or complex solutions, e.g. as in \cite{Maloney:2007ud, Banerjee:2009af, Alday:2012au}? When the dual theory is a TQFT in principle all observables are exactly computable in field theory, for many classes of theories defined on different conformal boundary manifolds. 
The AdS/CFT correspondence can then potentially help to clarify the answers to some of these questions, since the semi-classical 
gravity result must match the TQFT description.

Of course, one is tempted to push this line of argument further and speculate that this is a promising setting in which to try to formulate
a topological form of quantum gravity on the AdS side of the correspondence. Such a theory should be completely equivalent to the dual TQFT description. 
At present this looks challenging, to say the least, but there is an analogous construction in topological string theory. 
Here $U(N)$ Chern--Simons gauge theory (a Schwarz-type TQFT) on a three-manifold $M_3$ is equivalent to 
open topological strings on $T^*M_3$  \cite{Witten:1992fb}. There is a large $N$ duality relating this to a dual closed  topological string description.
For example, for $M_3=S^3$ the closed strings propagate on the resolved conifold background, with $N$ units of flux through the $S^2$ \cite{Ooguri:2002gx}. Here both sides are under computational control, 
and relate a TQFT to a topological sector of quantum gravity (string theory). This duality shares many features with AdS/CFT,\footnote{This was 
emphasized by C.~Vafa in his recent talk at the Princeton Workshop \emph{20 Years Later: The Many Faces of AdS/CFT}.} 
and might hint at how to attack the above problem.

In \cite{BenettiGenolini:2017zmu} we began much more modestly, setting up the basic problem in $\mathcal{N}=4$ gauged supergravity in five dimensions. With appropriate boundary conditions this defines the Donaldson--Witten topological twist of the dual $\mathcal{N}=2$ 
theory on the conformal boundary four-manifold, and we focused on the simplest observable, namely the partition function. 
Under AdS/CFT in the supergravity limit, minus the logarithm of the
partition function is identified with the holographically renormalized
supergravity action. We refer to this as the \emph{gravitational free energy} in this paper, and the main result of 
  \cite{BenettiGenolini:2017zmu} was to show that this is indeed a topological invariant, {\it i.e.}\ 
  it is invariant under arbitrary deformations of the boundary four-metric. The computation, although in principle straightforward, 
  was technically surprisingly involved. 
  Since 
  four-manifolds are also notoriously difficult, in this paper we set up an analogous problem in one dimension lower. 
The relevant bulk supergravity theory  is a Euclidean version of $\mathcal{N}=4$ $SO(4)$ gauged supergravity in four dimensions. 
  As well as the metric, the bosonic content of the theory contains two scalar fields and two  $SU(2)$ gauge fields. 
 Here $Spin(4)=SU(2)_{\CC}\times SU(2)_{\HH}$ is the spin double cover of $SO(4)$, and the fermions transform 
 in the fundamental $\mathbf{4}$ representation of this R-symmetry group. The topological twist 
 in particular identifies the boundary value of one of these two $SU(2)$ R-symmetry gauge fields with the spin connection of the conformal
 boundary three-manifold $(M_3,g)$. There is then a consistent truncation in which the other $SU(2)$ gauge field is identically zero in the bulk. Such Witten-type twists of $\mathcal{N}=4$ gauge theories in three dimensions have been studied in 
   \cite{Blau:1996bx}. 
      In the first part  of the paper we establish an analogous result to that in \cite{BenettiGenolini:2017zmu}, 
   namely that the gravitational free energy of such solutions is indeed invariant under arbitrary deformations of the boundary three-metric on $(M_3,g)$.
  
We  next analyse in more detail the geometry of supersymmetric solutions to the four-dimensional bulk supergravity theory. 
This geometry is characterized by what we call a \emph{twisted identity structure}. In an open set where the bulk spinor is 
non-chiral and the $SU(2)$ R-symmetry gauge bundle is trivialized, the spinor defines a canonical orthonormal frame $\{\mathrm{E}^a\}_{a=1,\ldots,4}$.
However, $\{\mathrm{E}^I\}_{I=1,2,3}$ rotate as the vector ${\bf 3}$ under the $SU(2)$ gauge group, while $\mathrm{E}^4$ 
is invariant, so that globally this frame is twisted by the R-symmetry bundle. We show that a supersymmetric solution to the bulk supergravity equations equivalently satisfies a certain 
first order differential system for this twisted identity structure. Using these equations, remarkably we are able to show that the bulk 
on-shell action is always a total derivative. By carefully analysing the global structure of the canonical twisted frame, and 
how this behaves where the bulk spinor becomes chiral or zero, this is shown to be globally a total derivative 
for any smooth solution. This is true on \emph{any} four-manifold $Y_4$ that fills any three-manifold boundary $M_3=\partial Y_4$. 
Moreover, on applying Stokes' theorem the bulk integral then always precisely cancels the boundary terms (including the holographic counterterms) in the action, with the net result being that the \emph{gravitational free energy of any smooth solution is zero}! 
Aided by this analysis, we return to the topological AdS$_5$/CFT$_4$ set up of \cite{BenettiGenolini:2017zmu}, 
and prove a precisely analogous result. Of course, here  not every four-manifold bounds a smooth five-manifold. 
 
At first sight these results are somewhat disappointing: the classical free energy is zero for smooth fillings, irrespective 
of their topology. Zero is a topological invariant, and not a very interesting one. However, if one believes that smooth real saddle points 
are the dominant saddle points in gravity, this is then a robust prediction for the large $N$ limits of various classes 
of topologically twisted SCFTs, in both three and four dimensions. For example, since $\mathcal{N}=4$ gauged supergravity in four dimensions \cite{Das:1977pu} 
is a consistent truncation of eleven-dimensional supergravity on $S^7$ (or $S^7/\Z_k$) \cite{Cvetic:1999au}, as we discuss later in the paper 
this leads to a prediction for the large $N$ limit of the partition function of the topologically twisted ABJM theory, on any three-manifold~$M_3$. 
On the other hand, with the exception of the $SU(N)$ Vafa--Witten partition function on $M_4=\mathrm{K3}$ discussed in section \ref{SecDiscussion},
to date none of these large $N$ limits have been computed in field theory: such computations now become 
very pressing! It might be that these match our supergravity results for smooth solutions, but if not then one necessarily has 
to consider more general saddle points, allowing e.g. for appropriate singularities and/or complex saddle points. Notice that 
although our computation of the classical gravitational free energy will in general break down for such solutions,
the result that this quantity is independent of boundary metric deformations is {\it a priori} a more general result. 
We have also so far only focused on the partition function, while in principle one should also be able to compute topological
correlation functions using similar holographic methods. We leave a fuller discussion of some of these issues to section \ref{SecDiscussion}.

The outline of the paper is as follows. First, in section \ref{Sec3dTFTs} we review the topological twists of three-dimensional supersymmetric field theories, as they are perhaps less well known than their four-dimensional relatives, and discuss the gravity dual to the ABJM 
theory. In section \ref{SecSUGRA}
we introduce the relevant four-dimensional $\mathcal{N}=4$ Euclidean gauged supergravity. Surprisingly 
the supersymmetry transformations of this theory, as formulated in \cite{Cvetic:1999au}, do not appear in the literature, and 
we hence first fill this gap. After holographically renormalizing the action, in section \ref{SecSUSY} 
we identify the conformal boundary Killing spinor equations 
which admit a topological twist 
as a particular solution on any oriented Riemannian three-manifold $(M_3,g)$. The bulk spinor equations are then 
expanded in a Fefferman--Graham-like expansion. In section \ref{SecVary} 
we prove that the gravitational free energy is independent of the metric $g$ on $M_3$, following a similar computation in 
\cite{BenettiGenolini:2017zmu}. In section \ref{SecGeometric} we show that a supersymmetric solution 
to the bulk equations of motion equivalently satisfies a first order differential system of equations 
for the twisted identity structure described above. Using this we prove that the 
gravitational free energy of any \emph{smooth} real solution is zero. In section \ref{SecRevisit}
we return to the AdS$_5$/CFT$_4$ correspondence in \cite{BenettiGenolini:2017zmu}, and 
prove an analogous result. We conclude in section \ref{SecDiscussion} with a more detailed discussion of some of the issues mentioned above.

\section{3d TQFTS and topological twists}\label{Sec3dTFTs}

We begin in section \ref{SecTwist} by reviewing topological twists of three-dimensional supersymmetric QFTs. In section \ref{SecABJM} 
we focus on the ABJM theory, its gravity dual, and the consistent truncation of eleven-dimensional supergravity 
on $S^7/\Z_k$ to four-dimensional $\mathcal{N}=4$ gauged supergravity.

\subsection{Twisting  $\mathcal{N}=4$ theories}\label{SecTwist}

One perspective on the topological twist is that it involves a modification of the global symmetry group of the theory, obtained by combining the spacetime symmetries with the R-symmetry of the theory. Concretely, one looks for group products such that a supercharge would transform as a singlet under an appropriate diagonal subgroup. In three dimensions every orientable manifold is spin.\footnote{This follows from the fact that in three dimensions the second Stiefel--Whitney class is the square of the first Stiefel--Whitney class, $w_2 = w_1^2$. Since a manifold is orientable if and only if $w_1=0$, we see that an orientable three-manifold is automatically spin.} Therefore, the frame bundle of any orientable three-manifold may be lifted to a $Spin(3)\cong SU(2)_E$, which constitutes the (Euclidean) spacetime symmetry.

\vspace{0.3cm}
On the other hand, the R-symmetry group of a three-dimensional field theory with $\mc{N}$ supersymmetries is $Spin(\mc{N})_R$. 
The minimal amount of supersymmetry required for a twist on a three-manifold with generic holonomy is $\mc{N}=4$: in the $\mc{N}=3$ case the supercharges transform as $(\mathbf{2},\mathbf{3})$ under $SU(2)_E\times Spin(3)_R$, and in the tensor product there is no singlet $\mathbf{2}\otimes\mathbf{3} = \mathbf{2}\oplus \mathbf{4}$. The R-symmetry group of $\mc{N}=4$ theories is $Spin(4)_R = SU(2)_{\CC}\times SU(2)_{\HH}$, and the supercharges transform as doublets under each of the two factors. The $\mc{N}=4$ multiplets are vector multiplets and hypermultiplets. The vector multiplet contains the gauge connection $\mathscr{A}$, a gaugino $\lambda$ and three real scalars $\vec{\phi}=(\phi_1,\phi_2,\phi_3)$, respectively transforming under $SU(2)_E\times SU(2)_{\CC}\times SU(2)_{\HH}$ as $(\mathbf{3},\mathbf{1},\mathbf{1})$, $(\mathbf{2},\mathbf{2},\mathbf{2})$ and $(\mathbf{1},\mathbf{3},\mathbf{1})$. The hypermultiplet contains two complex scalars $q$ and two spinors $\psi$, each forming R-symmetry doublets, that is, transforming as $(\mathbf{1},\mathbf{1},\mathbf{2})$ and $(\mathbf{2},\mathbf{2},\mathbf{1})$. There is an outer automorphism of the superalgebra exchanging $SU(2)_{\CC}$ and $SU(2)_{\HH}$. Under this automorphism, a vector multiplet is taken to a twisted vector multiplet, whose scalars transform as $(\mathbf{1},\mathbf{1},\mathbf{3})$, and a hypermultiplet is taken to a twisted hypermultiplet, whose scalars and spinors form doublets, respectively, of $SU(2)_{\CC}$ and $SU(2)_{\HH}$. The field components of the twisted multiplets will be denoted by a tilde.

One may twist using either $SU(2)_{\CC}$ or $SU(2)_{\HH}$, obtaining generically inequivalent TQFTs. The inequivalence of the two twists is not immediate from the supercharges: they transform as $(\mathbf{2},\mathbf{2},\mathbf{2})$ under $SU(2)_E\times SU(2)_{\CC}\times SU(2)_{\HH}$, so taking diagonal combinations of $SU(2)_E$ with either factors of the R-symmetry group leads to $(\mathbf{1},\mathbf{2})\oplus(\mathbf{3},\mathbf{2})$. Nevertheless, the twisted fields transform differently in the two twists, as can be seen from the scalars. For instance, consider the scalars in a hypermultiplet $q$: after the two twists, they would transform as $(\mathbf{1},\mathbf{2})$ under $(SU(2)_E\times SU(2)_{\CC})_{\rm diag}\times SU(2)_{\HH}$, or $(\mathbf{2},\mathbf{1})$ under $(SU(2)_E\times SU(2)_{\HH})_{\rm diag}\times SU(2)_{\CC}$. On the other hand, because of the exchange of $SU(2)_{\CC}$ and $SU(2)_{\HH}$, the scalars in the twisted hypermultiplet transform in the opposite way. The same goes for vector multiplets and twisted vector multiplets: the scalars in a vector multiplet form a triplet under $SU(2)_{\CC}$ and a singlet under $SU(2)_{\HH}$, so they distinguish between the two twists, but the opposite is true of the scalars in the twisted vector multiplet.

In a three-dimensional $\mc{N}=4$ super-Yang--Mills (SYM) theory, with a vector multiplet, the two twists are inequivalent. The first twist may also be recovered by dimensionally reducing the four-dimensional $\mc{N}=2$ Donaldson--Witten twist. The resulting model is sometimes referred to as super-BF or super-IG model, and the partition function reproduces the Casson invariant of the background three-manifold \cite{Witten:1989sx, Birmingham:1989is, Blau:1991bn}; and conjecturally, via renormalization group flow, the Rozansky--Witten invariants \cite{Blau:2000iy, Mikhaylov:2015nsa}.\footnote{More precisely, the Casson invariant arises when the gauge group $\mathscr{G}\cong SU(2)$, for three-manifolds $M_3$ with the same homology groups as $S^3$. It was originally defined in terms of the combinatorics of $SU(2)$-representations of $\pi_1(M_3)$. However, the Casson invariant naturally generalizes to the Lescop invariant, which is defined on any oriented three-manifold. Moreover, the TQFT Casson model suggests an extension of this invariant to any gauge group $\mathscr{G}$.} 
The second twist, instead, is intrisically three-dimensional (it is not known to arise from the reduction of any four-dimensional theory) and supposedly provides a mirror-symmetric description of the Casson invariant \cite{Blau:1996bx}. There exists a third topologically twisted three-dimensional SYM theory with two twisted scalar supercharges, which may be obtained by a partial twist of three-dimensional $\mc{N}=8$ SYM, or via dimensional reduction of the half-twist of four-dimensional $\mc{N}=4$ SYM. It is closely related to the Casson model, but differs from it by the matter content \cite{Geyer:2001yc}.

In three dimensions it is also possible to couple Chern--Simons theory to free hypermultiplets to obtain $\mc{N}=4$ supersymmetries \cite{Gaiotto:2008sd}, and twist the resulting theory \cite{Kapustin:2009cd, Koh:2009um}. As in the previous case, if there are only untwisted or twisted hypermultiplets in the theory the two twists are inequivalent, and usually referred to as an A-twist and B-twist, respectively. However, in a theory with both hypers and twisted hypers, the difference between the two twists amounts to the exchange between the untwisted and twisted matter. Therefore, one may consider a twist by a single factor in $Spin(4)_R$ and exchange the ``quality'' of the hypermultiplets, obtaining theories, sometimes called AB-models, which have both types of hypermultiplets. For concreteness, after the twist, an AB-model contains matter transforming under $(SU(2)_E\times SU(2)_{\CC})_{\rm diag}\times SU(2)_{\HH}$ as
\beq
\begin{split}
q \ &: \ (\mathbf{1}, \mathbf{1}, \mathbf{2}) \ \longmapsto \ (\mathbf{1},\mathbf{2})\, , \\
\psi \ &: \ (\mathbf{2}, \mathbf{2}, \mathbf{1}) \ \longmapsto \ (\mathbf{1},\mathbf{1})\oplus (\mathbf{3},\mathbf{1}) \, , \\
\tilde{q} \ &: \ (\mathbf{1}, \mathbf{2}, \mathbf{1}) \ \longmapsto \ (\mathbf{2},\mathbf{1})\, , \\
\tilde{\psi} \ &: \ (\mathbf{2}, \mathbf{1}, \mathbf{2}) \ \longmapsto \ (\mathbf{2},\mathbf{2})\, .
\end{split}
\eeq
Therefore, the bosonic fields are two scalars and a spinor, whilst the fermionic fields are a scalar, a one-form and two spinors. Chern--Simons-matter theories with $\mc{N}>4$ contain an equal number of untwisted and twisted hypermultiplets, so the symmetry between the A and B twist is automatically implemented.

In this paper we will be particularly interested in topological twists of the ABJM theory \cite{Aharony:2008ug} (see \cite{Lee:2008cr} for twists of the BLG 
\cite{Bagger:2006sk, Bagger:2007jr,  Bagger:2007vi, Gustavsson:2007vu} models).\footnote{The BLG models are Chern--Simons-matter theories with manifest $\mc{N}=8$ supersymmetry and concretely describe two M2-branes. On the other hand, ABJM theories, in the UV, are $\mc{N}=6$ $U(N)_k\times U(N)_{-k}$ Chern--Simons-matter theories describing $N$ M2-branes for any $N$. For $k=1,2$, the supersymmetry is enhanced to $\mc{N}=8$. For certain values of $N, k$ there exist equivalences between the BLG, ABJM and ABJ models \cite{Aharony:2008gk, Lambert:2010ji, Bashkirov:2011pt, Agmon:2017lga}.}
Classically this theory has $\mc{N}=6$ supersymmetry, so let us consider topological twists of $\mc{N}=6$ Chern--Simons-matter theories.
 Here the R-symmetry group is $Spin(6)_R\cong SU(4)$, and there are two decompositions
\beq
\begin{split}
(\mathrm{i}) \quad SU(4) \ & \longrightarrow \ SU(2)\times SU(2) \, , \\
(\mathrm{ii}) \quad SU(4) \ & \longrightarrow \ SU(2)\times SU(2)\times U(1) \, .
\end{split}
\eeq
In the first case we are viewing $SU(4)\cong Spin(6)$ as a double cover of $SO(6)\longmapsto SO(3)\times SO(3)$, the 
latter being the two diagonal $3\times 3$ blocks. In the second case instead the two copies of $SU(2)$ are the 
two diagonal $2\times 2$ blocks in $SU(4)$. Alternatively, projecting to $SO(6)$ the second decomposition 
is simply $SO(6)\longmapsto SO(4)\times SO(2)$, with the obvious $4+ 2$ block decomposition, 
where $SU(2)\times SU(2)\cong Spin(4)$ is the double cover of $SO(4)$, and $U(1)\cong SO(2)$.
The supercharges transform in the $\mathbf{6}$ of $SU(4)$, which decompose under the above as 
\beq
\begin{split}
(\mathrm{i}) \quad \mathbf{6} \ & \longrightarrow \ (\mathbf{1},\mathbf{3})\oplus (\mathbf{3},\mathbf{1}) \, , \\
(\mathrm{ii}) \quad \mathbf{6} \ & \longrightarrow \ (\mathbf{2},\mathbf{2})_{0} \oplus (\mathbf{1},\mathbf{1})_{+2} \oplus (\mathbf{1},\mathbf{1})_{-2} \, .
\end{split}
\eeq
In the first case it is clear that a twist with $SU(2)_E$ does not lead to any scalar supercharge, while for the second twist one reduces to the AB-model \cite{Koh:2009um}.

It is not completely  clear what the observables of the topologically twisted Chern--Simons-matter theories compute. In \cite{Koh:2009um} it was argued that the A-model is related via the novel Higgs mechanism \cite{Mukhi:2008ux} to the super-BF theory obtained by twisting $\mc{N}~=~4$ SYM, and thus computes the Casson invariant of the background three-manifold. Similarly, the mathematical content of the observables of the topological models of \cite{Kapustin:2009cd} is also currently unclear. 

The group-theoretic point of view on the topological twist considered above is not the only possible viewpoint. One may also describe the topological twist in the context of background rigid supersymmetry. For instance, in four dimensions the conditions for the background geometry to support a supersymmetric field theory have been studied by coupling to a non-dynamical supergravity \cite{Festuccia:2011ws}, and via holography \cite{Klare:2012gn}. In the first case it has been shown that the topological twist arises as a particular case where the $SU(2)$ connection corresponding to the gauged R-symmetry cancels part of the spin connection in the Killing spinor equation, thus allowing a scalar supercharge \cite{Karlhede:1988ax, Klare:2013dka}. In the second case it has been shown that the geometric structure of the bulk supergravity solutions reduces at the boundary to a quaternionic K\"ahler structure, which appears on any orientable Riemannian four-manifold \cite{BenettiGenolini:2017zmu}. Three-dimensional field theories with $\mc{N}=2$ have been extensively studied in the context of rigid supersymmetry, both from holography \cite{Klare:2012gn, Hristov:2013spa} and by coupling to supergravity \cite{Closset:2012ru}. However, the same cannot be said for $\mc{N}=4$ theories. As already mentioned, we will find very concretely that the topological twist corresponds to identifying the boundary value of one $SU(2)$ factor of the gauged R-symmetry with the spin connection. This allows us to construct a solution to the Killing spinor equation obtained from three-dimensional $\mc{N}=4$ conformal supergravity, in analogy with the standard approach. 

\subsection{The ABJM theory and its supergravity dual}\label{SecABJM}

The AdS/CFT correspondence has been especially influential in the context of three-dimensional field theories. 
In particular 
the AdS$_4\times S^7$ near-horizon geometry describing a stack of $N$ M2-branes provided strong evidence for the existence of a strongly-coupled maximally supersymmetric conformal field theory with $N^{3/2}$ degrees of freedom. After initial work by Bagger--Lambert--Gustavsson \cite{Bagger:2006sk, Bagger:2007jr,  Bagger:2007vi, Gustavsson:2007vu}, the worldvolume theory of $N$ M2-branes probing $\C^4/\Z_k$ was eventually found ten years ago by Aharony--Bergman--Jafferis--Maldacena \cite{Aharony:2008ug}.

The ABJM theory in flat spacetime $\mathbb{R}^{1,2}$  is conjectured to be holographically dual to M-theory on AdS$_4\times S^7/\Z_k$. In order to study the gravity dual of the field theory defined on different manifolds $M_3$ in the large $N$ limit, one may consider a consistent truncation of eleven-dimensional supergravity on $S^7$, or $S^7/\Z_k$, to an effective four-dimensional bulk supergravity theory. 
 Such a consistent truncation has been found 
in \cite{Cvetic:1999au}, where it is shown that any solution to the four-dimensional  $\mc{N}=4$  supergravity theory of Das--Fischler--Ro\v{c}ek \cite{Das:1977pu} uplifts to an eleven-dimensional solution. In particular this supergravity theory 
has a $Spin(4)\cong SU(2)\times SU(2)$ gauged R-symmetry, where the massless gauge fields arise, as usual 
in  Kaluza--Klein reduction, from a corresponding isometry of the internal space. Specifically, the uplifting/reduction 
ansatz in \cite{Cvetic:1999au} identifies the $SU(2)\times SU(2)$ isometry as acting in the ${\bf 2}$ of each factor in $\C^4\equiv \C^2\times \C^2$, 
where the internal space $S^7$ is the unit sphere in $\C^4$. This description makes it clear that one may 
also replace the internal space by $S^7/\Z_k$, where the $\Z_k$ acts on the coordinates of $\C^4$ via the diagonal action $z^i\mapsto \ex^{2\pi \ii/k}z^i$. This manifestly commutes with the $SU(2)\times SU(2)\subset SU(4)\curvearrowright\C^4$ action above. There is another notable geometric symmetry, namely the $\Z_2$ that acts by exchanging the two copies of $\C^2$ in $\C^4$, and thus exchanges the $SU(2)$ isometries. 
This symmetry is then inherited by the four-dimensional $\mc{N}=4$  gauged supergravity theory.

According to the holographic dictionary, symmetries of the eleven-dimensional solution correspond to symmetries of the field theory. 
In particular the $SU(2)\times SU(2)$ isometry of the internal space, which becomes a $Spin(4)_R$ gauged R-symmetry of 
the consistently truncated four-dimensional theory, corresponds to the $Spin(4)_R$ R-symmetry of the field theory dual.
The $\Z_2$ that acts as an outer automorphism, exchanging the  group factors in $Spin(4)_R\subset Spin(6)_R$, is indeed 
a symmetry of the $\mc{N}=6$ ABJM theory, since the latter 
has an equal number of untwisted and twisted hypermultiplets, in $\mc{N}=4$ language, and therefore its matter content is symmetric under the exchange of $SU(2)_{\CC}$ and $SU(2)_{\HH}$ \cite{Hosomichi:2008jb}. 

In the rest of the paper we will work entirely within the Das--Fischler--Ro\v{c}ek four-dimensional $\mc{N}=4$ gauged supergravity theory. 
Any solution to this theory, for a bulk asymptotically locally hyperbolic four-manifold $\B_4$, automatically uplifts on $S^7/\Z_k$ to give a gravity dual to the ABJM theory defined on the conformal boundary $M_3=\partial \B_4$. In particular we note that the 
effective four-dimensional Newton constant is
\begin{equation}\label{4dNewton}
\frac{1}{2\kappa_4^2} \ = \ \frac{k^{1/2}}{12\sqrt{2}\pi}N^{3/2}~.
\end{equation}

\section{Holographic supergravity theory}\label{SecSUGRA}

We begin in section \ref{SecDFR} by defining a real Euclidean section of $\mathcal{N}=4$  gauged supergravity in four dimensions and determine the fermionic supersymmetry transformations. A Fefferman--Graham expansion of asymptotically locally hyperbolic solutions to this theory is constructed in section \ref{SecFG}, for arbitrary conformal boundary three-manifold $(M_3,\met$). Using this, in section \ref{SecHoloRenormalization} we holographically renormalize the action.

\subsection{Euclidean $\mathcal{N}=4$ gauged supergravity}\label{SecDFR}

As outlined so far, holographic duals to three-dimensional SCFTs with a $Spin(4)=SU(2)_{\CC} \times SU(2)_{\HH}$ R-symmetry should be solutions of a four-dimensional $\mc{N}=4$ $SU(2)\times SU(2)$ gauged supergravity. 
As discussed in the previous subsection, 
the Das--Fischler--Ro\v{c}ek \cite{Das:1977pu}  theory  has a supersymmetric AdS$_4$ vacuum and was shown in \cite{Cvetic:1999au} to be a consistent truncation of eleven-dimensional supergravity on $S^7/\Z_k$.

In Lorentzian signature the bosonic sector of this $\mathcal{N}=4$ supergravity theory comprises the metric $\Met_{\mu\nu}$, two real scalars $\phi$, $\st$ which together parametrize an $SL(2,\mathbb{R})$ coset, and two triplets of $SU(2)$ gauge fields $\cA^I_\mu$, $\hat{\cA}^I_\mu$ ($I=1,2,3$). The associated field strengths are
\begin{align}
	\cF^I \ \equiv \ \dd \cA^I + \tfrac{1}{2} \gc\, \epsilon^{IJK} \cA^J \wedge \cA^K \, , \qquad \hat{\cF}^I \ \equiv \ \dd \hat{\cA}^I + \tfrac{1}{2} \gc\, \epsilon^{IJK} \hat{\cA}^J \wedge \hat{\cA}^K \, ,
\end{align}
and we have taken equal gauge couplings $\gc$ for each of the $SU(2)$ factors in the non-simple gauge group.
It is convenient to introduce the scalar field $X\equiv \ex^{\frac{1}{2} \phi}$ and define $\tilde{X} \equiv X^{-1} q$ where $q^2\equiv 1 + \st^2 X^4$.
The bosonic action and equations of motion in Lorentzian signature appear in \cite{Cvetic:1999au}.  However, as we are interested in holographic duals to TQFTs defined on Riemannian three-manifolds, we require a Euclidean signature version of this theory. After a Wick rotation the action becomes
\begin{align}
	I \ &= \ - \frac{1}{2 \kappa_4^2} \int \big[ R *1 - 2 X^{-2} \dd X \wedge * \dd X - \tfrac{1}{2} X^4 \dd \st \wedge * \dd \st + \gc^2 ( 8 + 2 X^2 + 2 \tilde{X}^2 ) * 1 \nn \\
	& \ \  - \tfrac{1}{2} X^{-2} \big( \cF^I \wedge * \cF^I + \ii \st X^2 \cF^I \wedge \cF^I \big) - \tfrac{1}{2} \tilde{X}^{-2} \big( \hat{\cF}^I \wedge * \hat{\cF}^I - \ii \st X^2 \hat{\cF}^I \wedge \hat{\cF}^I \big) \big] \, . \label{IEuclid}
\end{align}
Here $R=R(G)$ denotes the Ricci scalar of the metric $G_{\mu\nu}$, and $*$ is the Hodge duality operator acting on forms. The equations of motion which follow from this action are:\footnote{The Einstein equation \eqref{geom} incorporates the potential-like term which is missing from the Lorentzian version in \cite{Cvetic:1999au}.}
\begin{align}
	E_X : \ 0 \ =& \ \dd ( X^{-1} * \dd X ) - \tfrac{1}{2} X^4 \dd \st \wedge * \dd \st + \gc^2 \big( X^2 - X^{-2} ( 1 - \st^2 X^4 ) \big) *1 \label{Xeom} \\
	&\ + \tfrac{1}{4} X^{-2} \cF^I \wedge * \cF^I - \tfrac{1}{4} X^2 ( 1 - \st^2 X^4 ) q^{-4} \hat{\cF}^I \wedge * \hat{\cF}^I + \tfrac{\ii}{2} \st \tilde{X}^{-4} \hat{\cF}^I \wedge \hat{\cF}^I \, , \nn \\[5pt]
	E_\st : \ 0 \ =& \ \dd ( X^4 * \dd \st ) + 4 \gc^2 X^2 \st *1 - \tfrac{\ii}{2} \cF^I \wedge \cF^I \nn \\
	& \ + \st X^2 \tilde{X}^{-4} \hat{\cF}^I \wedge * \hat{\cF}^I + \tfrac{\ii}{2} ( 1 - \st^2 X^4 ) \tilde{X}^{-4} \hat{\cF}^I \wedge \hat{\cF}^I \, , \label{AxionEOM} \\[5pt]	
	E_{\cA^I} : \ 0 \ =& \ D ( X^{-2} * \cF^I ) + \ii \dd \st \wedge \cF^I \, , \label{AIeom} \\[5pt]
	E_{\hat{\cA}^I} : \ 0 \ =& \ \hat{D} ( \tilde{X}^{-2} * \hat{\cF}^I ) - \ii \dd ( \st X^2 \tilde{X}^{-2} ) \wedge \hat{\cF}^I \, , \label{hatAIeom} \\[5pt]	
	E_{G} : \ 0 \ =& \ R_{\mu\nu} + \gc^2 G_{\mu\nu} ( 4 + X^2 + \tilde{X}^2 ) - 2 X^{-2} \partial_\mu X \partial_\nu X - \tfrac{1}{2} X^4 \partial_\mu \st \partial_\nu \st \nn \\
	&\ - \tfrac{1}{2} X^{-2} \big( \cF^I_{\mu\rho} \cF^I_{\nu}{}^\rho - \tfrac{1}{4} G_{\mu\nu} ( \cF^I )^2 \big) - \tfrac{1}{2} \tilde{X}^{-2} \big( \hat{\cF}^I_{\mu\rho} \hat{\cF}^I_{\nu}{}^\rho - \tfrac{1}{4} G_{\mu\nu} ( \hat{\cF}^I )^2 \big) \label{geom} \, .
\end{align}
Here $( \cF^I )^2\equiv \sum_{I=1}^3\cF^I_{\mu\nu}\cF^{I\mu\nu}$, $( \hat{\cF}^I )^2\equiv \sum_{I=1}^3 \hat{\cF}^I_{\mu\nu} \hat{\cF}^{I\mu\nu}$ and the Bianchi identities define the $SU(2)$ covariant derivatives
\begin{align}
	B_{\cA^I}: \ D \cF^I \ \equiv & \ \dd \cF^I + \gc\, \epsilon^{IJK} \cA^J \wedge \cF^K \ = \ 0 \, , \\
	B_{\hat{\cA}^I}: \ \hat{D} \hat{\cF}^I \ \equiv & \ \dd \hat{\cF}^I + \gc\, \epsilon^{IJK} \hat{\cA}^J \wedge \hat{\cF}^K \ = \ 0 \, .
\end{align}
In general, equations \eqref{Xeom}--\eqref{geom} are complex, and solutions will likewise be complex. However, note that taking the axion $\st$ to be purely imaginary effectively removes all factors of $\ii$. Note also that the action and equations of motion are invariant under 
the $\Z_2$ symmetry: $\gc\rightarrow -\gc$, $\cA^I\rightarrow -\cA^I$, $\hat{\cA}^I\rightarrow -\hat{\cA}^I$. There is a second $\mathbb{Z}_2$ symmetry, discussed in section \ref{SecABJM}, which corresponds to the field theory outer automorphism exchanging the group factors in $Spin(4)_R\cong SU(2)_\CC\times SU(2)_\HH$. This second $\mathbb{Z}_2$ symmetry acts on the supergravity fields as $X \rightarrow \tilde{X}$, $\st X^2 \rightarrow - \st X^2$, $\cA^I \rightarrow \hat{\cA}^I$ and $\hat{\cA}^I \rightarrow \cA^I$. Whilst not manifest in the action and equations of motion, it can be made so upon rewriting the scalar kinetic terms in \eqref{IEuclid} as $2 X \tilde{X} \dd X \wedge * \dd \tilde{X} - \tfrac{1}{2} \dd ( \st X^2 ) \wedge * \dd ( \st X^2 )$.

In the Lorentzian theory the fermionic sector contains four gravitini, $\psi^a_\mu$, and four dilatini, $\chi^a$, which together with the spinor parameters $\epsilon^a$ all transform in the fundamental $\mathbf{4}$ representation of the $Spin(4)$ global R-symmetry group, which we label by $a=1,\ldots,4$. The supersymmetry transformations are not given in \cite{Cvetic:1999au} and the form of the action is different to that appearing in the original literature \cite{Das:1977pu}; in particular the parametrization of the scalars and their coupling to the gauge fields is different. We cannot, therefore, simply take the supersymmetry transformations given in \cite{Das:1977pu}. Of course, the two actions represent the same theory but presumably in different symplectic duality frames, and possibly with different gauge fixed $SL(2,\mathbb{R})$ scalar coset representatives. Instead of translating between the different presentations in Lorentzian signature and then Wick rotating to the Euclidean, we have instead derived the conditions for preserving supersymmetry by a different method. 

We started with a general ansatz for the gravitino and dilatino variations and then acted on the dilatino with the Dirac operator, adding additional field dependent multiples of the dilatino variation in order to recover a subset of the bosonic equations of motion \eqref{Xeom}--\eqref{geom}. This essentially shows that the dilatino field equation (in a bosonic background) maps to some of the bosonic field equations. Computing the integrability condition on the spinor parameter, which can be rephrased in terms of the free Rarita--Schwinger equation for the gravitino, and adding further dilatino variations recovers the remaining bosonic equations of motion. Hence the fermionic field equations map to bosonic ones, {\it i.e.}\ the theory is supersymmetric. At the end of this analysis we find:
\begin{align}
	\delta \psi_\mu^a \ = \ 0 \ =& \ \mathcal{D}_\mu \epsilon^a - \tfrac{1}{8\sqrt{2}} \eta^I_{ab} X^{-1} \cF^I_{\nu\lambda} \Gamma^{\nu\lambda} \Gamma_\mu \epsilon^b + \tfrac{1}{8\sqrt{2}} \bar{\eta}^I_{ab} X^{-1} \tilde{X}^{-2} \hat{\cF}^I_{\nu\lambda} \Gamma^{\nu\lambda} \Gamma_\mu \big[ 1 + \ii \st X^2 \Gamma_5 \big] \epsilon^b \nn \\
	& \ + \tfrac{\ii}{4} X^2 \partial_\mu \st \Gamma_5 \epsilon^a - \tfrac{1}{2\sqrt{2}} \gc \big[ ( X + X^{-1} )  - \ii \st X \Gamma_5 \big] \Gamma_\mu \epsilon^a \label{BulkGravitino} \, , \\[10pt]
	\delta \chi^a \ = \ 0 \ =& \ \tfrac{1}{8} \eta^I_{ab} X^{-1} \cF^I_{\nu\lambda} \Gamma^{\nu\lambda} \epsilon^b + \tfrac{1}{8} \bar{\eta}^I_{ab} X^{-1} \tilde{X}^{-2} \hat{\cF}^I_{\nu\lambda} \big[ 1 - \ii \st X^2 \Gamma_5 \big] \Gamma^{\nu\lambda} \epsilon^b \nn \\
	&\ + \tfrac{1}{\sqrt{2}} \big[ X^{-1} \partial_\nu X + \tfrac{\ii}{2} X^2 \partial_\nu \st \Gamma_5 \big] \Gamma^\nu \epsilon^a + \tfrac{1}{2} \gc \big[ ( X - X^{-1} ) + \ii \st X \Gamma_5 \big] \epsilon^a \label{BulkDilatino} \, .
\end{align}
Here the gauge covariant derivative acting on the supersymmetry parameter is
\begin{equation}
	\mathcal{D}_\mu \epsilon^a \ = \ \nabla_\mu \epsilon^a - \tfrac{1}{2} \gc\, \eta^I_{ab} \cA^I_\mu \epsilon^b + \tfrac{1}{2} \gc\, \bar{\eta}^I_{ab} \hat{\cA}^I_\mu \epsilon^b \, ,
\end{equation}
and $\eta^I_{ab}$, $\bar{\eta}^I_{ab}$ are respectively the self-dual/anti-self-dual 't Hooft symbols. In addition, $\Gamma_\mu$, $\mu=1,\ldots,4$, are generators of the Euclidean spacetime Clifford algebra, satisfying $\{\Gamma_\mu,\Gamma_\nu\}=2\Met_{\mu\nu}$, and we define $\Gamma_5 \equiv -\Gamma_{1234}$. Note that the $\Z_2$ symmetry that reverses the signs of $\gc$ and the two $SU(2)$ gauge fields is also a symmetry 
of these supersymmetry equations, provided one combines it with $\Gamma^\mu\rightarrow -\Gamma^\mu$.

For the purposes of completeness, we note that the transformations satisfy
\begin{align}
	\Gamma^\mu \mathcal{D}_\mu \delta \chi^a +& \tfrac{3\ii}{4} X^2 \partial_\mu \st \Gamma^\mu \Gamma_5 \delta \chi^a \nn \\
	=& \ \tfrac{1}{\sqrt{2}} E_X \epsilon^a - \tfrac{\ii}{2\sqrt{2}} X^{-2} E_\st \Gamma_5 \epsilon^a \nn \\
	&+ \tfrac{1}{8} \eta^I_{ab} X^{-1} ( B_{\cA^{I}} )_{\mu\nu\lambda} \Gamma^{\mu\nu\lambda} \epsilon^b + \tfrac{1}{8} \bar{\eta}^I_{ab} X^{-1} \tilde{X}^{-2} ( B_{\hat{\cA}^{I}} )_{\mu\nu\lambda} \Gamma^{\mu\nu\lambda} \big[ 1 - \ii \st X^2 \Gamma_5 \big] \epsilon^b \nn \\
	\label{eq:IntegrabilityDilatino}
	&+ \tfrac{1}{4} \eta^I_{ab} X ( E_{\cA^{I}} )_\mu \Gamma^\mu \epsilon^b + \tfrac{1}{4} \bar{\eta}^I_{ab} X^{-1} ( E_{\hat{\cA}^{I}} )_\mu \Gamma^\mu \big[ 1 - \ii \st X^2 \Gamma_5 \big] \epsilon^b \, , 
\end{align}
and
\begin{align}
	\Gamma^\nu [ \mathcal{D}_\mu , \mathcal{D}_\nu ] \epsilon^a &- \sqrt{2} X^{-1} \partial_\mu X \delta \chi^a + \tfrac{\ii}{\sqrt{2}} X^2 \partial_\mu \st \Gamma_5 \delta \chi^a - \tfrac{1}{2} \gc \big[ ( X - X^{-1} ) + \ii \st X \Gamma_5 \big] \Gamma_\mu \delta \chi^a \nn \\
	&+ \tfrac{1}{8} \eta^I_{ab} X^{-1} \cF^{I\nu\rho} \Gamma_{\nu\rho} \Gamma_\mu \delta \chi^b + \tfrac{1}{8} \bar{\eta}^I_{ab} X^{-1} \tilde{X}^{-2} \hat{\cF}^{I\nu\rho} \big[ 1 - \ii \st X^2 \Gamma_5 \big] \Gamma_{\nu\rho} \Gamma_\mu \delta \chi^b \nn \\
	=& \ \tfrac{1}{2} ( E_G )_{\mu\nu} \Gamma^\nu \epsilon^a - \tfrac{1}{8\sqrt{2}} \eta^I_{ab} X^{-1} ( B_{\cA^{I}} )^{\nu\rho\sigma} \Gamma_{\nu\rho\sigma} \Gamma_\mu \epsilon^b \nn \\
	&+ \tfrac{1}{8\sqrt{2}} \bar{\eta}^I_{ab} X^{-1} \tilde{X}^{-2} ( B_{\hat{\cA}^{I}} )^{\nu\rho\sigma} \Gamma_{\nu\rho\sigma} \Gamma_\mu \big[ 1 + \ii \st X^2 \Gamma_5 \big] \epsilon^b \nn \\	
	\label{eq:IntegrabilityGravitino}
	&- \tfrac{1}{4\sqrt{2}} \eta^I_{ab} X ( E_{\cA^{I}} )^\nu \Gamma_\nu \Gamma_\mu \epsilon^b + \tfrac{1}{4\sqrt{2}} \bar{\eta}^I_{ab} X^{-1} ( E_{\hat{\cA}^{I}} )^\nu \Gamma_\nu \Gamma_\mu [ 1 + \ii \st X^2 \Gamma_5 ] \epsilon^b \, .
\end{align}
In deriving these conditions we have not needed to specify the type of spinor we are using. Later, in section \ref{SecSUSY}, we will deal with a truncation of this theory in which one triplet of gauge fields is set to zero and the spinors are taken to be symplectic-Majorana.

\subsection{Fefferman--Graham expansion}\label{SecFG}

In this section we determine the Fefferman--Graham expansion \cite{Fefferman:2007rka} of asymptotically locally hyperbolic solutions to this Euclidean supergravity theory. This 
is the general solution to the bosonic equations of motion (\ref{Xeom})--(\ref{geom}), expressed as a perturbative expansion in a radial coordinate near the conformal boundary. 

We take the form of the metric to be \cite{Fefferman:2007rka}
\begin{align}
	G_{\mu\nu}\diff x^\mu \diff x^\nu \ = \ \frac{1}{z^2} \dd z^2 + \frac{1}{z^2} \mexp_{ij} \dd x^i\dd x^j \ = \ \frac{1}{z^2} \dd z^2 + h_{ij} \dd x^i\dd x^j \, . \label{FGmetric}
\end{align}
The AdS radius $\ell=1$, and in turn  we have the expansion
\begin{align}
	\mexp_{ij} \ = \ \mexp_{ij}^{0}+z^2 \mexp_{ij}^{2}+z^3 \mexp_{ij}^{3} + o(z^3) \, . \label{metricexp}
\end{align}
Here $\mexp_{ij}^{0}=\met_{ij}$ is the boundary metric induced on the conformal boundary $M_3$ at $z=0$. 

It is convenient to introduce the inner product $\langle \alpha , \beta \rangle$ between two $p$-forms $\alpha$, $\beta$ via
\begin{align}
	\alpha \wedge * \beta \ = \ \frac{1}{p!} \alpha_{\mu_1 \cdots \mu_p} \beta^{\mu_1 \cdots \mu_p}\,  \vol \, , %
\end{align}
where $\vol$ denotes the volume form, with associated Hodge duality operator $*$. 
The volume form for the four-dimensional bulk metric (\ref{FGmetric}) is 
\begin{align}
	\vol_4 \ =& \ \frac{1}{z^4}\dd z\wedge \vol_\mexp \ = \ \frac{1}{z^4}\dd z\wedge \sqrt{\det \mexp} \, \dd x^1\wedge \dd x^2 \wedge \dd x^3 \, .\label{volform}
\end{align}
The determinant may then be expanded in a series in $z$, around that for $\mexp^0$, as follows
\begin{align}
\sqrt{\det \mexp} \ =  \sqrt{\det \mexp^0} \, \Big[ & 1 + \tfrac{z^2}{2} t^{(2)} + \tfrac{z^3}{2} t^{(3)} \Big] + o(z^3) \, .
\end{align}
Here we have denoted $t^{(n)} \equiv \mathrm{Tr} \left[ (\mexp^0)^{-1} \mexp^n \right]$ and indices are always raised with $\mexp^0$.

The remaining bosonic fields are likewise expanded as follows:
\begin{align}
	X \ =& \ 1 + z X_1 + z^2 X_2 + z^3 X_3 + o(z^3)\, , \label{Xexp}  \\[5pt]
	\st \ =& \ z \st_1 + z^2 \st_2 + z^3 \st_3 + o(z^3) \, , \\[5pt]
	\cA^I \ =& \ A^I + z \ma_1^I + z^2 \ma_2^I + o(z^2) \, , \label{aIexp} \\[5pt]
	\hat{\cA}^I \ =& \ \hat{A}^I + z \hat{\ma}_1^I + z^2 \hat{\ma}_2^I + o(z^2) \, . \label{hataIexp}
\end{align}
We have chosen a gauge in which all $\dd z$ terms in the gauge field expansions are set to zero.
 
We now substitute the above expansions into the equations of motion \eqref{Xeom}--\eqref{geom} and solve them order by order in the radial coordinate $z$ in terms of the boundary data $\mexp^0 = g, \Xb, \st_1$, $A^I$ and $\hat{A}^I$. For the Einstein equation (\ref{geom}) we will need the Ricci tensor of the metric \eqref{FGmetric}:
\begin{align}
	R_{zz} \ =& \ - \frac{3}{z^2}-\frac{1}{2}\Big(\mathrm{Tr}\left[\mexp^{-1}\partial^2_z\mexp\right]-\tfrac{1}{z}\mathrm{Tr}\left[\mexp^{-1}\partial_z\mexp\right]-\tfrac{1}{2}\mathrm{Tr}\left[\mexp^{-1}\partial_z\mexp\right]^2\Big) \, ,\label{Rzz}\\[5pt]
	R_{ij} \ =& \ -\frac{3}{z^2}\mexp_{ij}-\Big(\tfrac{1}{2}\partial^2_z\mexp-\tfrac{1}{z}\partial_z\mexp-\tfrac{1}{2}(\partial_z\mexp)\mexp^{-1}(\partial_z\mexp)+\tfrac{1}{4}(\partial_z\mexp)\mathrm{Tr}\left[\mexp^{-1}\partial_z\mexp\right]\nn\\
	&\hspace{2.8cm} - R(\mexp)-\tfrac{1}{2z}\mexp\mathrm{Tr}\left[\mexp^{-1}\partial_z\mexp\right]\Big)_{ij}\, ,\label{Rij}\\[5pt]
	R_{zi} \ =& \ - \frac{1}{2}(\mexp^{-1})^{jk}\Big(\nabla_i\mexp_{jk,z}-\nabla_k\mexp_{ij,z}\Big)\, ,\label{Rzi}
\end{align}
where $\nabla$ is the covariant derivative for $\mexp$. 

Examining first the axion equation (\ref{AxionEOM}) gives at the first two orders
\begin{equation}
	0 \ = \ ( 1 - 2 \gc^2 ) \st_1 \, , \qquad 0 \ = \ ( 1 - 2 \gc^2 ) ( 2 X_1 \st_1 + \st_2 ) \, ,
\end{equation}
which can be solved by setting $\gc = \pm \frac{1}{\sqrt{2}}$. These equations fix the gauging coupling in terms of the AdS$_4$ length scale, which we have set to unity. At even higher order we find
\begin{align}
	\nabla^2 \st_1  \ =& \ 2 \gc^2 \Big( \st_1 ( t^{(2)} + 2 X_1^2 + 4 X_2 ) + 4 X_1 \st_2 + 2 \st_3 \Big) \, . \label{LapAxion1}
\end{align}

Moving on to the dilaton equation \eqref{Xeom} we find
\begin{align}
	0 \ = \ ( 1 - 2 \gc^2 ) X_1 \, , \qquad 0 \ = \ ( 1 - 2 \gc^2 )( X_2 - \tfrac{1}{2} X_1^2 + \tfrac{1}{4} \st_1^2 ) \, ,
\end{align}
which are again solved by $\gc=\pm \frac{1}{\sqrt{2}}$ together with 
\begin{align}
	\nabla^2 X_1 \ =& \ 2 \gc^2 \Big( 2 X_3 + X_1 ( t^{(2)} + 2 X_1^2 - 2 X_2 + \st_1^2 ) + \st_1 \st_2 \Big) - 2 \st_1 ( X_1 \st_1 + \st_2 )  \, . \label{LapDilaton1}
\end{align}

Next the $\cA^I$ gauge field equation (\ref{AIeom}) yields 
\begin{align}
	0 \ = \ D *_{\mexp^0} \ma_1^I \, , \qquad \ma_2^I \ = \ X_1 \ma_1^I + \tfrac{1}{2} *_{\mexp^0} D *_{\mexp^0} F^I - \tfrac{\ii}{2} \st_1 *_{\mexp^0} F^I \label{aI2eqn} \, , 
	\end{align}
where the curvature is $F^I \equiv \dd A^I + \tfrac{1}{2} \gc\, \epsilon^{IJK} A^J \wedge A^K$. Notice that $\ma_1^I$, and hence $\ma_2^I$, is partially undetermined. Similarly, the other gauge field equation (\ref{hatAIeom}) gives
\begin{align}
	0 \ = \ \hat{D} *_{\mexp^0} \hat{\ma}_1^I \, , \qquad \hat{\ma}_2^I \ = \ - X_1 \hat{\ma}_1^I + \tfrac{1}{2} *_{\mexp^0} \hat{D} *_{\mexp^0} \hat{F}^I + \tfrac{\ii}{2} \st_1 *_{\mexp^0} \hat{F}^I \label{hataI2eqn} \, , 
	\end{align}
with $\hat{F}^I \equiv \dd \hat{A}^I + \tfrac{1}{2} \gc\, \epsilon^{IJK} \hat{A}^J \wedge \hat{A}^K$.

The non-trivial information from the $ij$ component of the Einstein equation \eqref{geom}, using \eqref{Rij}, is
\begin{align}
	\mexp^2_{ij} \ =& \ - \big[ R_{ij}(\mexp^0) - \tfrac{1}{4} \mexp^0_{ij} R(\mexp^0) \big] - \mexp^0_{ij} \big( \tfrac{1}{2} X_1^2 + \tfrac{1}{8} \st_1^2 \big) \, , \label{g2}
\end{align}
which is a matter-modified version of the boundary Schouten tensor.
From this expression we immediately deduce that the trace of $\mexp^2_{ij}$ is
\begin{align}
	t^{(2)} \ =& \ - \tfrac{1}{4} R(\mexp^0) - \tfrac{3}{2} X_1^2 - \tfrac{3}{8} \st_1^2 \, .
\end{align}
The $zz$ component of the Einstein equation in \eqref{geom}, together with \eqref{Rzz}, determines the trace of the highest order component in the expansion of the bulk metric:
\begin{align}
	t^{(3)} \ = \ \tfrac{4}{3} X_1^3 - \tfrac{2}{3} X_1 ( 4 X_2 + \st_1^2 ) - \tfrac{2}{3} \st_1 \st_2 \, .
\end{align}

\subsection{Holographic renormalization}\label{SecHoloRenormalization}

Having solved the bulk equations of motion to the relevant order, we are now in a position to holographically renormalize the Euclidean $\mathcal{N}=4$  gauged supergravity theory. The bulk action~(\ref{IEuclid}) is divergent for an asymptotically locally hyperbolic solution, but can be rendered finite by the addition of appropriate local counterterms. We begin by taking the trace of the Einstein equation \eqref{geom}. Substituting the result into the Euclidean action \eqref{IEuclid} with $\gc=\pm \frac{1}{\sqrt{2}}$, we arrive at the bulk on-shell action 
\begin{align}
	I_{\text{o-s}} \ = \ \frac{1}{2\kappa_4^2} \int_{\B_4} \ \Big[ &- ( 4 + X^2 + \tilde{X}^2 ) * 1 - \tfrac{1}{2} X^{-2} \big( \cF^I \wedge * \cF^I + \ii \st X^2 \cF^I \wedge \cF^I \big) \nn \\
	&- \tfrac{1}{2} \tilde{X}^{-2} \big( \hat{\cF}^I \wedge * \hat{\cF}^I - \ii \st X^2 \hat{\cF}^I \wedge \hat{\cF}^I \big)\Big]\, . \label{IEuclidOnShell}
\end{align}
Here $\B_4$ is the bulk four-manifold, with boundary $\partial \B_4=M_3$. In order to obtain the equations of motion \eqref{Xeom}--\eqref{geom} from the original bulk action \eqref{IEuclid} on a manifold with boundary, one has to add the Gibbons--Hawking--York term
\begin{align} \label{eq:GHY}
	I_{\text{GHY}} \ =& \ - \frac{1}{\kappa_4^2} \int_{\partial \B_4} \dd^3 x \, \sqrt{\det h} \, K \ = \ \frac{1}{\kappa_4^2} \int_{\partial \B_4} \dd^3 x \, {z} \partial_z \sqrt{\det h} \, .
\end{align}
Here more precisely one cuts $\B_4$ off at some finite radial distance, or equivalently non-zero $z>0$, 
and $(M_3,h)$ is the resulting three-manifold boundary, with trace of the second fundamental form being $K$. 
Recall from (\ref{FGmetric}) that $h_{ij} = \frac{1}{z^2} \mexp_{ij}$.

The combined action $I_{\text{o-s}}+I_{\text{GHY}}$ suffers from divergences as the conformal boundary is approached. To remove these divergences we use the standard method of holographic  renormalization \cite{Emparan:1999pm, Taylor:2000xw, deHaro:2000vlm}. Namely, we introduce a small cut-off $z=\cutoff>0$, and expand all fields via the Fefferman--Graham expansion of section \ref{SecFG} to identify the divergences. 
These may be cancelled by adding local boundary counterterms. We find
\begin{align}
	I_{\text{ct}} \ = \ 
	\frac{1}{\kappa_4^2} \int_{\partial \B_4} &\dd^3 x \, \sqrt{\det h}\, \Big[  {2} +\tfrac{1}{2} R ( h ) + ( X - 1 )^2 + \tfrac{1}{4} \st^2 \Big] \, .\label{ctaction}
\end{align}
As is standard, we have written the counterterm action (\ref{ctaction}) covariantly in terms of the induced metric $h_{ij}$ on $M_3=\partial \B_4$.
The total renormalized action is then
\beq
\Sren \ = \ \lim_{\cutoff\rightarrow 0} \, \left(I_{\text{o-s}} + I_{\text{GHY}}+I_{\text{ct}}\right)\, ,\label{Actionfinal}
\eeq
which by construction is finite.

The choice of counterterms (\ref{ctaction}) defines a particular renormalization scheme. 
For this theory there are other local, gauge invariant counterterms that one can construct from the boundary fields, 
that have non-zero (and finite) limits as $\cutoff \rightarrow 0$. It is straightforward to check that there are no such 
finite counterterms constructed without using the scalar fields; but including the latter we may write down finite counterterms 
proportional to the boundary integrals of $\st^3$, $(X-1)^3$, $\st R(h)$, {\it etc}. There are also  
local but non-gauge invariant terms that one might consider. For example, boundary Chern--Simons terms for the $SU(2)$ gauge fields, 
and the boundary gravitational Chern--Simons term. However, such terms would change the gauge invariance of the theory, 
and  we shall hence not consider them further.\footnote{The topological twist will later identify one boundary $SU(2)$ gauge field with the 
boundary spin connection of $(M_3,g)$, so that these Chern--Simons terms are the same. Moreover, since any oriented 
three-manifold is parallelizable there is always a globally defined frame. Choosing such a frame then allows one to interpret the 
gravitational Chern--Simons term as a global three-form on $M_3$. However, its integral  depends on the choice of framing.} In principle we should use a \emph{supersymmetric} holographic renormalization scheme, 
but in the absence of a prescription for this we shall use the minimal scheme with counterterms (\ref{ctaction}) in the remainder of the paper, 
cf. the discussion in \cite{Genolini:2016sxe, Genolini:2016ecx, Papadimitriou:2017kzw, An:2017ihs}.
In any case, for the topological twist boundary condition the boundary values $\st_1$, $X_1$ of $\st$ and $X$ will be zero, and the above-mentioned 
finite gauge invariant counterterms are all zero.

Given the renormalized action we may compute the following vacuum expectation values (VEVs):
\begin{align}
	\langle T_{ij} \rangle \ =& \ \frac{2}{\sqrt{g}} \frac{\delta \Sren}{ \delta g^{ij} } \, , \, \, \qquad \langle \Xi \rangle \ = \ \frac{1}{\sqrt{g}} \frac{\delta \Sren}{ \delta \Xb } \, , \qquad \langle \Sigma \rangle \ = \ \frac{1}{\sqrt{g}} \frac{\delta \Sren}{ \delta \st_1 } \, , \nn\\ 
	\langle \Jcur_I^{i} \rangle \ = & \ \frac{1}{\sqrt{g}} \frac{\delta \Sren}{ \delta A^I_i } \, , 
	\qquad \langle \hat{\Jcur}_I^{i} \rangle \ = \ \frac{1}{\sqrt{g}} \frac{\delta \Sren}{ \delta \hat{A}^I_i } \, .
	\label{VEVs}
\end{align}
Here, as usual in AdS/CFT, the boundary fields $\met_{ij}$, $\Xb$, $\st_1$, $A^I_i$ and $\hat{A}^I_i$ act as sources for operators, and 
the expressions in (\ref{VEVs}) compute the VEVs of these operators. Using 
the above holographic renormalization we may write (\ref{VEVs}) as the following limits:
\begin{align}
	\langle T_{ij} \rangle \ =& \ \frac{1}{\kappa_4^2} \lim_{\cutoff \to 0} \, \frac{1}{\cutoff} \Big[ - K_{ij} + K h_{ij} +  R_{ij}(h) - \tfrac{1}{2} h_{ij} R(h)  + h_{ij} ( - 2 - ( X - 1 )^2 - \tfrac{1}{4} \st^2 ) \Big] \, , \nn \\[10pt]
	\langle \Xi \rangle \ =& \ \frac{1}{\kappa_4^2} \lim_{\cutoff \to 0} \, \frac{1}{\cutoff^2} \Big[ - 2 \cutoff X^{-2} \partial_\cutoff X + 2 ( X - 1 ) \Big] \, ,\nn \\[10pt]
	\langle \Sigma \rangle \ =& \ \frac{1}{\kappa_4^2} \lim_{\cutoff \rightarrow 0} \frac{1}{\cutoff^2} \Big[ - \tfrac{1}{2} \cutoff X^4 \partial_\cutoff \st + \tfrac{1}{2} \st \Big] \, , \nn \\[10pt]
	\langle \Jcur^{Ii} \rangle \ =& \ \frac{1}{2\kappa_4^2} \lim_{\cutoff \to 0} \, \frac{1}{\cutoff^3} \Big[- *_h \Big( \dd x^i \wedge \big( X^{-2} *_4 \cF^I +  \ii \st \cF^I \big) \Big) \Big] \, ,\nn\\[10pt]
	\langle \hat{\Jcur}^{Ii} \rangle \ =& \ \frac{1}{2\kappa_4^2} \lim_{\cutoff \to 0} \, \frac{1}{\cutoff^3} \Big[- *_h \Big( \dd x^i \wedge \big( \tilde{X}^{-2} *_4 \hat{\cF}^I - \ii \st X^2 \tilde{X}^{-2} \hat{\cF}^I \big) \Big] \, . 
\end{align}
Here $K_{ij}$ is the second fundamental form of the cut-off hypersurface $(M_3,h_{ij}$), and $*_h$ denotes the Hodge duality operator for the  metric $h_{ij}$. A computation then gives the finite expressions
\begin{align}
	\langle T_{ij} \rangle \ =& \ \ \frac{1}{\kappa_4^2} \Big[ \tfrac{3}{2} \mexp^3_{ij} - \tfrac{1}{2} \mexp^0_{ij} \big( 3 t^{(3)} + 4 X_1 X_2 + \st_1 \st_2 \big) \Big]  \, ,\label{Tij} \\[10pt]
	\langle \Xi \rangle \ =& \ \ \frac{1}{\kappa_4^2} \big( 4 X_1^2 - 2 X_2 \big) \, , \label{Xi}\\[10pt]
	\langle \Sigma \rangle \ =& \ - \frac{1}{\kappa_4^2} \big( 2 X_1 \st_1 + \tfrac{1}{2} \st_2 \big) \, , \label{Sigma}\\[10pt]	
	\langle \Jcur^{I}_i \rangle \ =& \ -\frac{1}{2\kappa_4^2}(\ma_1^I)_i\label{calJi}~,\\[10pt]
	\langle \hat{\Jcur}^{I}_i \rangle \ =& \ -\frac{1}{2\kappa_4^2}(\hat{\ma}^I_1)_i\label{bbJ} \, .
\end{align}
Notice that each of these expressions contain terms that are not determined, in terms of boundary data, by the Fefferman--Graham expansion of the bosonic equations of motion. In particular the $\mexp^3_{ij}$ term in the stress-energy tensor $T_{ij}$, the scalars $\Xs$, $\st_2$ that determine respectively $\Xi$, $\Sigma$, and $\ma_1^I$, $\hat{\ma}_1^I$ appearing in the $SU(2)_R$ currents. 

As a quick check/application of these formulae, consider a boundary Weyl transformation $\delta \sigma$ under which $\delta g^{ij} = 2 g^{ij} \delta \sigma$, the scalars $X_1, \st_1$ have Weyl weight 1: $\delta X_1 = X_1 \delta \sigma$, $\delta \st_1 = \st_1 \delta \sigma$ and the gauge fields Weyl weight 0. Then it is a simple exercise to show that 
\begin{align}
	\delta_\sigma \Sren \ = \ \int_{\partial \B_4} \vol_g \Big[ \tfrac{1}{2} T_{ij} \delta g^{ij} + \Xi \delta X_1 + \Sigma \delta \st_1 + \Jcur^{I}_i \delta A^{Ii} + \hat{\Jcur}^{I}_i \delta \hat{A}^{Ii} \Big] \ = \ 0 \, , 
\end{align}
which is consistent with the fact that there is no conformal anomaly in three-dimensional SCFTs.


\section{Supersymmetric solutions}\label{SecSUSY}

In this section we study supersymmetric solutions to the Euclidean $\mathcal{N}=4$ supergravity theory. We begin in section \ref{SecBoundaryKSE} by deriving the Killing spinor equations on the conformal boundary from the bulk supersymmetry equations, and then compare them to the component form equations of off-shell three-dimensional $\mathcal{N}=4$ conformal supergravity. In section \ref{SecTT} we describe how the topological twist arises as a special solution to these Killing spinor equations, that exists on any Riemannian three-manifold $(M_3,g)$. Finally, in section \ref{SecSUSYexpand} we expand solutions to the bulk spinor equations in a Fefferman--Graham-like expansion.

\subsection{Boundary spinor equations}\label{SecBoundaryKSE}

We begin by introducing the charge conjugation matrix $\mathscr{C}$ for the Euclidean spacetime Clifford algebra. By definition $\Gamma_\mu^*=\mathscr{C}^{-1}\Gamma_\mu \mathscr{C}$, and one may choose Hermitian generators $\Gamma_\mu^\dagger=\Gamma_\mu$ together with the conditions $\mathscr{C}=\mathscr{C}^*=-\mathscr{C}^{\mathrm{T}}$, $\mathscr{C}^2=-1$. We may then define spinors in Euclidean signature to satisfy the symplectic-Majorana condition
\beq\label{Bulkcc}
\epsilon^a \ \equiv \ \Omega^a{}_b \mathscr{C}(\epsilon^b)^*~,
\eeq
with $\Omega = \sigma_3 \otimes \ii \sigma_2$.
It is straightforward to check that when $\hat{\mc{A}}^I=0$, and provided the axion $\st$ is purely imaginary with all other bosonic fields being real, the supersymmetry variations (\ref{BulkGravitino}), (\ref{BulkDilatino}) are compatible with this symplectic-Majorana condition. We will be interested in solutions that satisfy these reality conditions, and henceforth work in the truncation of the bulk supergravity theory for which the triplet of $SU(2)$ gauge fields $\hat{\cA}^I_\mu$ is set to zero.  For completeness we record here the truncated bulk supersymmetry conditions:
\begin{align}
	0 \ =& \ \nabla_\mu \epsilon^a - \tfrac{1}{2} \gc \eta^I_{ab} \cA^I_\mu \epsilon^b - \tfrac{1}{8\sqrt{2}} \eta^I_{ab} X^{-1} \cF^I_{\nu\lambda} \Gamma^{\nu\lambda} \Gamma_\mu \epsilon^b + \tfrac{\ii}{4} X^2 \partial_\mu \st \Gamma_5 \epsilon^a \nn \\
	&\ - \tfrac{1}{2\sqrt{2}} \gc \big[ ( X + X^{-1} ) - \ii \st X \Gamma_5 \big] \Gamma_\mu \epsilon^a \label{BulkGravitinoAHatzero} \, , \\[10pt]
	 0 \ =& \ \tfrac{1}{8} \eta^I_{ab} X^{-1} \cF^I_{\nu\lambda} \Gamma^{\nu\lambda} \epsilon^b + \tfrac{1}{\sqrt{2}} \big[ X^{-1} \partial_\nu X + \tfrac{\ii}{2} X^2 \partial_\nu \st \Gamma_5 \big] \Gamma^\nu \epsilon^a \nn \\
	 &\ + \tfrac{1}{2} \gc \big[ ( X - X^{-1} ) + \ii \st X \Gamma_5 \big] \epsilon^a \label{BulkDilatinoAHatzero} \, .
\end{align}

We next expand the bulk Killing spinor equations \eqref{BulkGravitinoAHatzero}, 
\eqref{BulkDilatinoAHatzero} to leading order near the conformal boundary at $z=0$. We will consequently need the 
Fefferman--Graham expansion of an orthonormal frame for the metric (\ref{FGmetric}), (\ref{metricexp}), together with the associated spin connection. The following is a choice of frame ${\rm E}_\mu^{\overline{\mu}}$ for the metric (\ref{FGmetric}):
\begin{equation}
{\rm E}^{\overline{z}}_z \ = \ \frac{1}{z}, \qquad {\rm E}^{\overline{z}}_i \ = \ {\rm E}^{\overline{i}}_z \ =  \ 0, \qquad  {\rm E}_i^{\overline{i}} \ = \ \frac{1}{z}\te^{\overline{i}}_i~,\label{FGframe}
\end{equation}
where $\te^{\overline{i}}_i$ is a frame for the $z$-dependent metric $\mexp$. The latter then has the expansion (\ref{metricexp}), but for the 
present subsection we shall only need that
\begin{equation}
\te^{\overline{i}}_i \ = \ \ex^{\overline{i}}_i + O(z^2) \, ,
\end{equation}
where $ \ex^{\overline{i}}_i$ is a frame for the boundary metric $\mexp^0=g$.
The non-zero components of the  spin connection $\Omega_\mu^{\ \overline{\nu}\overline{\rho}}$ at this order 
are correspondingly
\begin{align}
 \Omega_{{i}}^{\ \overline{zj}} \ = \ \frac{1}{z}\ex_{i}^{\, \overline{j}}+O(z)\, , \qquad \Omega_{{i}}^{\ \overline{j k}} \ =\  \omega_{{i}}^{\ \overline{jk}} + O(z^2) \, ,
\end{align}
where $\omega_{{i}}^{\ \overline{j k}}$ denotes the boundary spin connection. 

We take as the generators of the Clifford algebra the following 
\begin{align}
	\Gamma_{\bar{1}} \ \equiv \ \Gamma_{\bar{z}} \ = \ \left(\begin{array}{cc} \identity_2 & 0 \\ 0 & - \identity_2 \end{array}\right) \, , \quad \Gamma_{\overline{1+i}} \ = \ \left(\begin{array}{cc} 0 & \sigma_{\bar{i}} \\ \sigma_{\bar{i}} & 0 \end{array}\right) \,  ,
\end{align}
so that 
\begin{equation}
 \Gamma_5 \ = \ \left(\begin{array}{cc} 0 & -\ii \identity_2 \\ \ii \identity_2 & 0 \end{array}\right)~,\label{Gamma5}
\end{equation}
and 
\beq
\mathscr{C} \ = \ \begin{pmatrix}
\ii \sigma_2 & 0 \\
0 & - \ii \sigma_2
\end{pmatrix} ~,
\eeq
where $\sigma_{\bar{i}}$ the usual Pauli matrices. The bulk Killing spinor is then expanded as
\begin{equation}
	\epsilon^a \ = \ z^{-1/2} \los^a + z^{1/2} \xi^a + o(z^{1/2})\, .
\end{equation}
From the $z$-component of the gravitino equation (\ref{BulkGravitinoAHatzero}) one then finds
\begin{align}
	0 \ =& \ - z^{-1/2} \tfrac{1}{2} ( \identity \pm \Gamma_{\bar{z}} ) \los^a + z^{1/2} \left[ \tfrac{1}{2} ( \identity \mp \Gamma_{\bar{z}} ) \xi^a + \tfrac{\ii}{4} \st_1 \Gamma_5 ( \identity \pm \Gamma_{\bar{z}} ) \los^a \right] + o(z^{1/2}) \, ,
\end{align}
with the upper/lower signs corresponding to taking $\gc = \pm \frac{1}{\sqrt{2}}$. We can then satisfy this equation by taking $\los^a$ to have a definite chirality under $\Gamma_{\bar{z}}$ and $\xi^a$ to have the opposite chirality. Recall that there is a $\Z_2$ symmetry of the action, equations 
of motion, and supersymmetry equations, that sends $\gc\rightarrow-\gc$, $\cA^I\rightarrow -\cA^I$, $\Gamma^\mu\rightarrow -\Gamma^\mu$. 
Using this, without loss of generality we set 
 $\gc = - \frac{1}{\sqrt{2}}$ from now on, so that $\los^a$ has positive $\Gamma_{\bar{z}}$ chirality and $\xi^a$ negative chirality, and we write them as
\begin{align}
	\los^a \ = \ \left(\begin{array}{cc} \los^a_L \\ 0 \end{array}\right) \, , \qquad \xi^a \ = \ \left(\begin{array}{cc} 0 \\ \xi^a_R \end{array}\right) \, .\label{Leadingspinors}
\end{align}

The leading order term in the $i$-component of the gravitino equation is then seen to be identically satisfied. The next order gives the boundary Killing spinor equation (KSE):
\begin{align}
	0 \ =& \ \nabla^A_{{i}} \los^a_L + \sigma_{i} \xi^a_R - \tfrac{1}{4} \st_1 \sigma_{i} \los^a_L \, . \label{bdyKSE}
\end{align}
Here $\nabla^A_{i} \los^a_L = \nabla_{{i}} \los^a_L + \tfrac{1}{2\sqrt{2}} \eta^I_{ab} A^I_i \los^b_L$, where the covariant derivative is with respect to the Levi--Civita spin connection of the boundary metric $\mexp^0_{ij}=g_{ij}$, and $\sigma_i = \sigma_{\bar{i}}\, \ex^{\bar{i}}_i$, so that $\{\sigma_i,\sigma_j \}=2g_{ij}$. Note that after redefining the conformal spinor parameter such that $\tilde{\xi}^a_R = \xi^a_R - \frac{1}{4} \st_1 \los^a_L$, the boundary KSE becomes
\begin{align}
	0 \ =& \ \nabla^A_{{i}} \los^a_L + \sigma_{i} \tilde{\xi}^a_R \, . \label{confKSE}
\end{align}
This is the equation which results from setting to zero the gravitino supersymmetry variation of off-shell 3d $ \mc{N}=4$ conformal supergravity \cite{Banerjee:2015uee}.

Turning to the bulk dilatino equation \eqref{BulkDilatinoAHatzero}, the leading order term is equivalent to the chirality property of $\los^a$. At the next order we obtain two conditions, corresponding to the left and right-handed components
\begin{align}
	0 \ =& \ - \tfrac{1}{\sqrt{2}} \st_1 \xi^a_R - \tfrac{1}{2\sqrt{2}} ( X_1^2 - 2 X_2 ) \los^a_L + \tfrac{1}{2\sqrt{2}} \partial_i \st_1 \sigma^i \los^a_L + \tfrac{1}{8} \eta^I_{ab} F_{ij}^I \sigma^{ij} \los^b_L \, , \label{lodilu} \\[10pt]
	0 \ =& \ \sqrt{2} X_1 \xi^a_R + \tfrac{1}{2\sqrt{2}} ( X_1 \st_1 + \st_2 ) \los^a_L - \tfrac{1}{\sqrt{2}} \partial_i X_1 \sigma^i \los^a_L + \tfrac{1}{4} \eta^I_{ab} ( \ma_1^I )_i \sigma^i \los^b_L \label{lodild} \, .
\end{align}
After the redefinition of the conformal spinor parameter and Hodge dualising one term these read
\begin{align}
	0 \ =& \ - \tfrac{1}{\sqrt{2}} \st_1 \tilde{\xi}^a_R - \tfrac{1}{2\sqrt{2}} \big( \tfrac{1}{2} \st_1^2 + X_1^2 - 2 X_2 \big) \los^a_L + \tfrac{1}{2\sqrt{2}} \partial_i \st_1 \sigma^i \los^a_L + \tfrac{1}{8} \eta^I_{ab} F_{ij}^I \sigma^{ij} \los^b_L \, , \label{lodiludef} \\[10pt]
	0 \ =& \ \sqrt{2} X_1 \tilde{\xi}^a_R + \tfrac{1}{\sqrt{2}} \big( X_1 \st_1 + \tfrac{1}{2} \st_2 \big) \los^a_L - \tfrac{1}{\sqrt{2}} \partial_i X_1 \sigma^i \los^a_L - \tfrac{\ii}{8} \eta^I_{ab} ( * \ma_1^I )_{ij} \sigma^{ij} \los^b_L \label{lodilddef} \, .
\end{align}
These equations are not equivalent, and matching them to the single algebraic condition arising from setting a three-dimensional dilatino variation to zero is not therefore entirely straightforward. The Weyl multiplet of off-shell $\mc{N}=4$ conformal supergravity contains two auxiliary scalar fields $S_1$, $S_2$ of Weyl weight 1 and 2 respectively, and  generically six gauge fields. The vanishing of the dilatino supersymmetry transformation \cite{Banerjee:2015uee} when one triplet of gauge fields is turned off is, schematically, 
\begin{align}
	0 \ =& \ S_1 \tilde{\xi}^a + S_2 \los^a + \partial_i S_1 \sigma^i \los^a + \eta^I_{ab} F^I_{ij} \sigma^{ij} \los^b \, . \label{N44dconfdil}
\end{align}
Clearly \eqref{lodiludef} is of this form once we identify $S_1 \sim \st_1$, $S_2 \sim \tfrac{1}{2} \st_1^2 + X_1^2 - 2 X_2$. However, \eqref{lodilddef} does not match so neatly as $* \ma_1^I$ is not a field strength. Moreover, our spinor expansion should recover a single equation, and so it is perhaps some linear combination of \eqref{lodiludef} and \eqref{lodilddef} that reproduces \eqref{N44dconfdil}. In any case, it is not clear that 
the leading order dilatino equation should match this particular off-shell formulation of $\mc{N}=4$ conformal supergravity.

\subsection{Topological twist}\label{SecTT}

Recall that the boundary Killing spinor equation \eqref{bdyKSE} written in full is
\begin{align}
	0 \ = \ \partial_i \los^a_L + \tfrac{1}{4} \omega_i{}^{\overline{jk}} \sigma_{\overline{jk}} \los^a_L + \tfrac{1}{2\sqrt{2}} \eta^I_{ab} A^I_i \los^b_L + \sigma_{i} \xi^a_R - \tfrac{1}{4} \st_1 \sigma_{i} \los^a_L \, .
\end{align}
To solve this equation with a topological twist, we begin by setting the boundary scalar $\st_1$ and conformal spinor parameter $\xi^a_R$ to zero. We then identify the boundary $SU(2)$ gauge field with the spin connection as follows
\begin{align}
	A^I_i \ = \ \tfrac{1}{\sqrt{2}} \epsilon^I{}_{\overline{jk}} \omega_{i}{}^{\overline{jk}} \, . \label{twistcond}
\end{align}
The constant spinor which solves the Killing spinor equation is then
\begin{equation}
	\los^a_{L} \ = \ \ii\sigma^a \begin{pmatrix} w \\ \ii\bar{w} \end{pmatrix} \, , \label{twistsoln}
\end{equation}
where $w$ is any complex number and 
\begin{align}
	(\sigma^a) = (\sigma^1,\sigma^2,\sigma^3,-\ii \identity_2) \, . \label{extPauli}
\end{align}
It is useful to note that the 't Hooft symbol action on $\los^a_L$ may be exchanged for the Pauli matrix action:
\begin{align}
	\eta^I_{ab} \los^b_L \ =& \ - \ii \sigma^I \los^a_L \, . \label{proj}
\end{align}

We have solved the leading order KSE. Turning to the algebraic spinor equations we note that, in general, the conformal spinor parameter $\xi^a_R$ can be solved for by taking the $\sigma^i$ trace of the KSE \eqref{bdyKSE}. Substituting this generic expression for $\xi^a_R$ into \eqref{lodilu} and rescaling by $\sqrt{2}$ leads to
\begin{align}
	0 \ =& \ - \st_1 \sla{\nabla}^A \los^a_L + \tfrac{1}{2} [\sla{\nabla}^A , \sla{\nabla}^A ] \los^a_L + \tfrac{1}{2} \partial_i \st_1 \sigma^i \los^a_L + \tfrac{1}{4} ( 3 \st_1^2 - 2 X_1^2 + 4 X_2 + R ) \los^a_L \, , \label{bdyDil}
\end{align}
with $R=R(g)$ the boundary Ricci scalar. Specialising to the field configuration which solves the boundary KSE above, this simplifies to
\begin{align}
	0 \ =& \ \tfrac{1}{4} ( - 2 X_1^2 + 4 X_2 + R ) \los^a_L \, , 
\end{align}
and therefore fixes 
\begin{align}
	X_2 \ = \ \tfrac{1}{4} ( 2 X_1^2 - R ) \, .
\end{align}

The other algebraic relation \eqref{lodild} now reads
\begin{align}
	0 \ =& \ \tfrac{1}{2\sqrt{2}} \st_2 \los^a_L - \tfrac{1}{\sqrt{2}} \partial_i X_1 \sigma^i \los^a_L + \tfrac{1}{4} \eta^I_{ab} ( \ma_1^I )_i \sigma^i \los^b_L \, .
\end{align}
Here recall that $\ma_1^I$ is (proportional to) the VEV of the remaining $SU(2)_R$ current. One can use \eqref{proj} to swap the 't Hooft symbol for a Pauli matrix, plus the usual relation
\begin{align}
	\sigma_{\bar{i}}\sigma_{\bar{j}} \ = \ \delta_{\overline{ij}} + \ii \epsilon_{\overline{ijk}}\sigma_{\bar{k}}~.
\end{align}
The resulting equation takes the algebraic form
\begin{align}
	c_b \sigma^b \los^a_L \ =& \ 0~, \label{algebraiclemma}
\end{align}
where $(\sigma^b)$ are the extended Pauli matrices \eqref{extPauli}, and the coefficients $c_b$ are \emph{real}. In particular here we use that $\varphi_2$ is purely imaginary. Using the solution \eqref{twistsoln}, one can easily check that as long as $w\neq 0$ equation \eqref{algebraiclemma} implies that $c_a=0$ for all $a=1,2,3,4$. We thus conclude the equations
\begin{align}
	\varphi_2 \ =& \ \tfrac{\ii}{\sqrt{2}}(\ma_1^I)_{\bar{i}}\, \delta^{\bar{i}}_I \, , \qquad \partial_{\bar{i}} X_1 \ = \ \tfrac{1}{2\sqrt{2}} \epsilon_{\overline{ij}I} (\ma_1^I)^{\bar{j}} \, . \label{vphi2}
\end{align}
Note here the trace over frame indices and $SU(2)_R$ indices in the expression for $\varphi_2$: this makes sense globally, since the topological twist identifies the gauge bundle with the spin bundle. Having identified indices we may view $(\ma_1^I)^{\bar{i}}$ as a two-tensor.

\subsection{Supersymmetric expansion}\label{SecSUSYexpand}

In this section we continue to expand the bulk spinor equations to higher order in $z$. From this we extract further information about some of the fields which are not fixed, in terms of boundary data, by the bosonic equations of motion. 
We will continue to use the boundary conditions appropriate to the topological twist. The frame, spin connection and spinor expansions beyond the leading order given in section \ref{SecBoundaryKSE} will be needed, so we first give details of these. The frame expansion is 
\beq
\label{eq:FrameExpansion}
\mathtt{e}^{\overline{i}}_i \ = \ \ex^{\overline{i}}_i + \tfrac{1}{2} z^2 ( \mexp^2 )^{\overline{i}}{}_{\overline{j}} \, \ex^{\overline{j}}_i+ z^3 (\ex^{(3)})^{\overline{i}}_i + o(z^3) \, ,
\eeq
where in particular $\ex^{\overline{i}}_i $ is a frame for the boundary metric and we have used a local SO(3) rotation to gauge fix the order $z^2$ term. The additional spin connection components we will need are
\begin{align}
\label{eq:SpinConnection5d}
	\Omega_i{}^{\overline{zi}} \ =& \ \frac{1}{z}\mathtt{e}^{\overline{i}}_i - \tfrac{1}{2} \mexp^{jk} \mathtt{e}^{\overline{i}}_j\partial_z\mexp_{ik}\, , \qquad
	\Omega_z{}^{\overline{ij}} \ = \ \ \mexp^{ij}\mathtt{e}^{[\overline{i}}_i \partial_z\mathtt{e}^{\overline{j}]}_j \, .
\end{align}
The bulk spinor then has the following expansion
\begin{equation}
	\epsilon^a \ =  \ z^{-1/2} \los^a + z^{3/2}\varepsilon^a_3 + z^{5/2} \varepsilon^a_5 + o(z^{5/2})\, ,\label{genspinorexp}
\end{equation}
where $\los^a$ are constant with positive chirality under $\Gamma_{\bar{z}}$. 

The remaining orders of the bulk dilatino equation give us
\begin{align}
	0 \ =& \ \tfrac{1}{2\sqrt{2}} ( X_1^3 - 4 X_1 X_2 + 4 X_3 ) \los^a_L + \tfrac{1}{2\sqrt{2}} \partial_{\bar{i}} \st_2 \sigma^{\bar{i}} \los^a_L + \tfrac{1}{8} \eta^I_{ab} \big( ( F^I_1 )_{\overline{ij}} - X_1 F^I_{\overline{ij}} \big) \sigma^{\overline{ij}} \los^b_L \label{Dil52u} \, , \\[10pt]
	0 \ =& \ - \sqrt{2} X_1 \varepsilon^a_{3,R} - \tfrac{1}{2\sqrt{2}} ( 3 X_1 \st_2 + 2 \st_3 ) \los^a_L + \tfrac{1}{\sqrt{2}} \big( \partial_{\bar{i}} X_2 - X_1 \partial_{\bar{i}} X_1 \big) \sigma^{\bar{i}} \los^a_L \nn \\
	& \ - \tfrac{1}{4} \eta^I_{ab} \big( 2( \ma_2^I )_{\bar{i}} - X_1 ( \ma_1^I )_{\bar{i}} \big) \sigma^{\bar{i}} \los^b_L \label{Dil52l} \, ,
\end{align}
where $F_1^I = D \ma^I_1 \equiv \dd \ma^I_1 - \tfrac{1}{\sqrt{2}} \epsilon^{IJK} A^J \wedge \ma^K_1$. The remaining gravitino expansions give
\begin{align}
	0 \ =& \ \varepsilon^a_{3,L} + \tfrac{1}{8} X_1^2 \los^a_L - \tfrac{1}{16\sqrt{2}} \eta^I_{ab} F^I_{\overline{ij}} \sigma^{\overline{ij}} \varepsilon^b_L \label{Gravzo32u} \, , \\[10pt]
	0 \ =& \ \varepsilon^a_{3,R} - \tfrac{1}{4} \st_2 \varepsilon^a_L + \tfrac{1}{4\sqrt{2}} \eta^I_{ab} ( \ma_1^I )_{\bar{i}} \sigma^{\bar{i}} \los^b_L \label{Gravzo32l} \, , \\[10pt]
	0 \ =& \ \tfrac{1}{2} \mexp^2_{\overline{ij}} \sigma^{\bar{j}} \varepsilon^a_L + \tfrac{1}{4} X_1^2 \sigma_{\bar{i}}\los^a_L - \tfrac{1}{8\sqrt{2}} \eta^I_{ab} F^I_{\overline{jk}} \sigma^{\overline{jk}} \sigma_{\bar{i}} \los^b_L \label{Gravza32l} \, , \\[10pt]
	0 \ =& \ \varepsilon^a_{5,L} - \tfrac{1}{12} ( X_1^3 - 2 X_1 X_2 ) \los^a_L - \tfrac{1}{24\sqrt{2}} \eta^I_{ab} \big( ( F_1^I )_{\overline{ij}} - X_1 F^I_{\overline{ij}} \big) \sigma^{\overline{ij}} \los^b_L \label{Gravzo52u} \, , \\[10pt]
	0 \ =& \ \varepsilon^a_{5,R} - \tfrac{1}{8} ( 3 X_1 \st_2 + 2 \st_3 ) \los^a_L + \tfrac{1}{8\sqrt{2}} \eta^I_{ab} \big( 2 ( \ma^I_2 )_{\bar{i}} - X_1 ( \ma^I_1 )_{\bar{i}} \big) \sigma^{\bar{i}} \los^b_L \label{Gravzo52l} \, , \\[10pt]
	0 \ =& \ \sigma_{\bar{i}} \varepsilon^a_{5,R} + \nabla^A_{\bar{i}} \varepsilon^a_{3,L} + \tfrac{1}{4} \omega^{(2) \overline{jk}}_{\bar{i}} \sigma_{\overline{jk}} \los^a_L - \tfrac{1}{4} ( X_1 \st_2 + \st_3 ) \sigma_{\bar{i}} \los^a_L \nn \\
	& \ - \tfrac{1}{4\sqrt{2}} \eta^I_{ab} \big( ( \mexp^2 )_{\bar{i}}{}^{\bar{j}} A^I_{\bar{j}} - X_1 ( \ma^I_1 )_{\bar{i}} \big) \varepsilon^b_L + \tfrac{1}{4\sqrt{2}} \eta^I_{ab} \big( 2 ( \ma^I_2 )_{\bar{j}} - X_1 ( \ma^I_1 )_{\bar{j}} \big) \sigma^{\overline{ij}} \varepsilon^b_L \label{Gravza52u} \, , \\[10pt]	
	0 \ =& \ \tfrac{3}{4} \mexp^3_{\overline{ij}} \sigma^{\bar{j}} \los^a_L + \nabla^A_{\bar{i}} \varepsilon^a_{3,R} - \tfrac{1}{4} ( X_1^3 - 2 X_1 X_2 ) \sigma_{\bar{i}} \los^a_L - \tfrac{1}{4} \partial_{\bar{i}} \st_2 \los^a_L \nn \\
	&\ - \tfrac{1}{8\sqrt{2}} \eta^I_{ab} \big( ( F^I_1 )_{\overline{jk}} - X_1 F^I_{\overline{jk}} \big) \sigma^{\overline{jk}} \sigma_{\bar{i}} \varepsilon^b_L \label{Gravza52l} \, .
\end{align}
From the topological twist condition \eqref{twistcond} the boundary gauge field strength is 
\begin{align}
	F^I_{\overline{ij}} \ = \ \tfrac{1}{\sqrt{2}} \epsilon^I{}_{\overline{kl}} R_{\overline{ij}}{}^{\overline{kl}} \, .
\end{align}
Substituting this and the expressions for $X_2$, $\ma^I_1$ and $\st_2$ into \eqref{Gravzo32u}, \eqref{Gravzo32l} allows us to identify 
\begin{align}
	\varepsilon^a_{3,L} \ = \ - \tfrac{1}{16} ( 2 X_1^2 - R ) \, \los^a_L \, , \qquad \varepsilon^a_{3,R} \ = \  \tfrac{1}{2} \st_2 \, \los^a_L  - \tfrac{1}{2} \partial_{\bar{i}} X_1 \sigma^{\bar{i}} \los^a_L \, . \label{varepsilon3}
\end{align}
We also find that equation \eqref{Gravza32l} is identically satisfied given the expression \eqref{g2} for $\mexp^2$ found in solving the Einstein equation. Equations \eqref{Gravzo52u} and \eqref{Gravzo52l} are  solved by removing the unknown quantities $F^I_1$, $\ma^I_2$ using \eqref{Dil52u} and \eqref{Dil52l}:  
\begin{align}
	\varepsilon^a_{5,L} \ =& \ - \tfrac{1}{24} ( X_1 R - 2 X_1^3 + 8 X_3 ) \los^a_L - \tfrac{1}{12} \partial_{\bar{i}} \st_2 \sigma^{\bar{i}} \, \los^a_L \, , \\[10pt]
	\varepsilon^a_{5,R} \ =& \ \tfrac{1}{2} ( 2 X_1 \st_2 + \st_3 ) \los^a_L - \tfrac{1}{16} \partial_{\bar{i}} ( 2 X_1^2 - R ) \, \sigma^{\bar{i}} \los^a_L \, .
\end{align} 
We will not solve \eqref{Gravza52u} as knowledge of $\ma^I_2$ or $\omega^{(2)}$ is not relevant for our purposes. Turning now to \eqref{Gravza52l}, using previous results we can re-express this particular equation as 
\begin{align}
	0 \ =& \ \big[ \tfrac{3}{4} \mexp^3_{\overline{ij}} - \tfrac{1}{2} \nabla_{\bar{i}} \partial_{\bar{j}} X_1 - \tfrac{1}{8} X_1 R \delta_{\overline{ij}} \big] \sigma^{\bar{j}} \los^a_L \nn \\
	& \ + \tfrac{1}{4} \partial_{\bar{i}} \st_2 \los^a_L - \tfrac{1}{8\sqrt{2}} \eta^I_{ab} \big( ( F^I_1 )_{\overline{jk}} - X_1 F^I_{\overline{jk}} \big) \sigma^{\overline{jk}} \sigma_{\bar{i}} \varepsilon^b_L\, .
\end{align}  
By taking the real part we can extract the remaining term in the Fefferman--Graham expansion of the bulk metric 
\begin{align}
	\mexp^3_{\overline{ij}} =& \ \tfrac{2}{3} \nabla_{\overline{i}}\partial_{\overline{j}}X_1 + \tfrac{1}{6} X_1 R\delta_{\overline{ij}} + \tfrac{1}{6\sqrt{2}} (F_{1(\overline{i}})^{\overline{kl}}\epsilon_{\overline{j})\overline{kl}} - \tfrac{1}{3\sqrt{2}}(F_1^{\overline{k}})^{\overline{l}}_{\ph{l}( \overline{i}}\epsilon_{\overline{j})\overline{kl}} \nn \\
	& \ - X_1 \big[ \tfrac{1}{6\sqrt{2}} (F_{(\overline{i}})^{\overline{kl}}\epsilon_{\overline{j})\overline{kl}} - \tfrac{1}{3\sqrt{2}} (F^{\overline{k}})^{\overline{l}}_{\ph{l}( \overline{i}}\epsilon_{\overline{j})\overline{kl}} \big] \label{g3} \, .
\end{align}


\section{Metric independence}\label{SecVary}

Our aim in this short section is to show that, for any supersymmetric asymptotically locally hyperbolic solution 
to the Euclidean $\mathcal{N}=4$ supergravity theory, with the topologically twisted boundary conditions on 
an arbitrary Riemannian three-manifold $(M_3,g)$, 
the variation with respect to the arbitrary boundary metric of the holographically renormalized action is identically zero.  

An arbitrary deformation of the renormalized action can be written as
\begin{align}
	\delta \Sren \ = \ \int_{\partial \B_4=M_3} \dd^3 x \, \sqrt{\det g} \Big[ \tfrac{1}{2} T_{ij} \delta g^{ij} + \Xi \delta X_1 + \Sigma \delta \st_1 + \Jcur^{I}_i \delta A^{Ii} + \hat{\Jcur}^{I}_i \delta \hat{A}^{Ii} \Big] \, .\label{Srenvary}
\end{align}
For the topological twist we set $\st_1=0$ and $A^I_i = \frac{1}{\sqrt{2}} \epsilon^I{}_{\overline{jk}} \omega_{i}{}^{\overline{jk}}$, together with truncating the bulk $SU(2)$ triplet $\hat{\cA}^{I}=0$. At this point we have not chosen a value for the freely specifiable boundary field $X_1$ which, recall, has Weyl weight 1. In order for $\delta X_1$ to be relatable to $\delta g^{ij}$, $X_1$ must be a scalar function built from the boundary curvature tensors, $R_{ijkl}, R_{ij}$ and $R$. However, from these tensors we cannot construct a Weyl weight 1 object. Consequently we choose to set $X_1=0$ as part of the topological twist boundary conditions. 

To evaluate $\delta A^I_i$ we require the variation of the boundary spin connection in terms of the boundary metric:
\begin{align}
	\delta \omega_{i}{}^{\overline{jk}} \ = \ \tfrac{1}{2} \ex^{j\bar{j}} \ex^{k\bar{k}} ( \nabla_k \delta g_{ij} - \nabla_j \delta g_{ik} ) \, .
\end{align}
Thus
\begin{align}
	\delta A^I_i \ = \ \tfrac{1}{\sqrt{2}} \epsilon^I{}_{\overline{jk}} \delta \omega_{i}{}^{\overline{jk}} \ = \ \tfrac{1}{\sqrt{2}} \epsilon^I{}_{\overline{jk}} \ex^{j\bar{j}} \ex^{k\bar{k}} \nabla_k \delta g_{ij} \, .
\end{align}
Therefore the variation of the action for the topological twist boundary conditions reduces to
\begin{align}
	\delta \Sren = \int_{M_3} &\left[ \left( \tfrac{1}{2} T_{ij} - \tfrac{1}{\sqrt{2}} \nabla^k ( \Jcur_{Ii} \epsilon^I{}_{\overline{jk}} \ex_j^{\bar{j}} \ex_k^{\bar{k}} ) \right) \delta g^{ij} + \nabla_k \left( \tfrac{1}{\sqrt{2}} \epsilon^I{}_{\overline{jk}} \, \Jcur_I^i \ex^{j\bar{j}} \ex^{k\bar{k}} \delta g_{ij}  \right) \right] \vol_3 \, ,
\end{align}
where we have introduced $\vol_3\equiv \sqrt{ \det g}\, \diff^3 x$.
Dropping the total derivative, which is zero for the closed three-manifolds we are considering, and inserting the expressions for the stress-energy tensor and $SU(2)$ current from \eqref{Tij} and \eqref{calJi} gives
\begin{align}
	\delta \Sren \ = \ \frac{1}{4\kappa_4^2} \int_{M_3} \mc{T}_{ij} \delta g^{ij} \vol_3 \, , 
\end{align}
where the effective stress-energy tensor is
\begin{align}
	\mc{T}_{ij} \ =& \ 3 \mexp^3_{ij} + \tfrac{1}{\sqrt{2}} \nabla^k \big( \epsilon_{Ik (i} \, ( \ma^I_1 )_{j)} \big) \, .
\end{align}
Note that because we have identified spacetime and R-symmetry indices, the covariant derivative in $\mc{T}_{ij}$ acts on both the $I$ and $i$ indices of $( \ma^I_1 )_{i}$. Inserting the expression for $\mexp^3_{ij}$ from \eqref{g3} when $X_1=0$ gives 
\begin{align}
	\mc{T}_{ij} \ =& \ \ex^{\bar{i}}_i \ex^{\bar{j}}_j \big[ \tfrac{1}{2\sqrt{2}} (F_{1(\overline{i}})^{\overline{kl}}\epsilon_{\overline{j})\overline{kl}} - \tfrac{1}{\sqrt{2}}(F_1^{\overline{k}})^{\overline{l}}_{\ph{l}( \overline{i}}\epsilon_{\overline{j})\overline{kl}} \big] + \tfrac{1}{\sqrt{2}} \nabla^k \big( \epsilon_{Ik (i} ( \ma^I_1 )_{j)} \big)~.
\end{align}
Expanding the field strengths we have
\begin{align}
	2 \sqrt{2} \mc{T}_{ij} \ =& \ \ex^{\bar{i}}_i \ex^{\bar{j}}_j \big[ \nabla^{\overline{k}}(\ma_{1(\overline{i}})^{\overline{l}}\epsilon_{\overline{j})\overline{kl}} + (\omega^{\overline{k}})_{(\overline{i}}{}^{I} (\ma_{1|I|})^{\overline{l}} \epsilon_{\overline{j})\overline{kl}} + 2\nabla_{[\overline{l}}(\ma^{\overline{k}}_1)_{(\overline{i}]} \epsilon_{\overline{j})}{}^{\overline{l}}{}_{\overline{k}} + 2(\omega_{[\overline{l}})^{\overline{k}I}(\ma_{1|I|})_{(\overline{i}]} \epsilon_{\overline{j})}{}^{\overline{l}}{}_{\overline{k}} \big]\nn \\
	&+ 2 \nabla^k \big( \epsilon_{Ik (i} ( \ma^I_1 )_{j)} \big) \, .
\end{align}
Here covariant derivatives of $( \ma^I_1 )_i$ in the first line are understood to act with respect to the index outside the bracket only, in contrast to the action on the second line. By carefully expanding, using the definition of the spin connection as the connection of the frame bundle, and recalling from section \ref{SecTT} that when $X_1=0$, $( \ma^I_1 )_{i}$ is symmetric in $I$ and $i$ indices, we find delicate cancellations and ultimately that $\mc{T}_{ij} =0$. Notice this is true for an arbitrary background closed  three-manifold $(M_3,g)$, and 
that while the Fefferman--Graham expansion does not determine $( \ma^I_1 )_i$, nevertheless the expression for $\mc{T}_{ij}$ is identically 
zero.

We close this section by commenting on more precisely when the derivation in this section holds, and in particular when the formula 
(\ref{Srenvary}) holds. The latter computes the variation $\delta \Sren$ of the on-shell action. A variation of the boundary fields induces a corresponding 
variation of the bulk fields. Since the background solution that we are varying about solves the 
bulk equations of motion, crucially the bulk contribution to the resulting variation of the on-shell action is zero (by definition, 
this bulk integrand multiplies the bulk equations of motion). Thus $\delta \Sren$ is necessarily a boundary term, and for smooth saddle point 
solutions dual to the vacuum, one expects the only boundary to be the conformal boundary $\partial \B_4=M_3$. Equation 
(\ref{Srenvary}) is the resulting boundary expression. However, this computation would also hold if the bulk solution is singular, or 
has internal boundaries, provided these do not contribute a corresponding surface term in the interior, in addition to (\ref{Srenvary}). 
The internal boundary conditions for fields are clearly then relevant, but if one is going to allow internal singularities/boundaries 
of this type in a putative saddle point, the absence of these additional surface terms is a fairly clear constraint. 

\section{Geometric reformulation}\label{SecGeometric}

In this section we first reformulate the bulk supersymmetry conditions \eqref{BulkGravitinoAHatzero}, \eqref{BulkDilatinoAHatzero} in terms of a local identity structure. We then use this structure in section \ref{Sec3dOS} to determine the renormalized on-shell action for \emph{any} smooth filling with topological twist boundary conditions.

\subsection{Twisted identity structure}\label{SecId}

Recall that the bulk spinor is originally a quadruplet of Dirac spinors, and we halved the number of degrees of freedom by requiring that it solve the symplectic-Majorana condition \eqref{Bulkcc}. Therefore, the quadruplet of spinors has the form
\begin{align}
	\epsilon^a \ =& \ \left( \epsilon^1 \, , - (\epsilon^1)^c \, , \epsilon^2 \, , (\epsilon^2)^c \right)^{\mathrm{T}} \, ,
\end{align}
where $\epsilon^{1,2}$ are Dirac spinors on the four-manifold $\B_4$ and the charge conjugate is $\epsilon^c = \mathscr{C}\epsilon^*$. Notice that the Weyl condition imposed with $\Gamma_5$ acting on the spinor indices is not compatible with the topological twist. One sees this from 
the expressions (\ref{Gamma5}) and (\ref{Leadingspinors}): the leading order term in the expansion of the bulk spinor is 
chiral if and only if it is zero.
However, we may instead act with $\Gamma_5$ on the R-symmetry indices of the spinor and require
\beq\label{eq:InternalChiralityEq}
(\Gamma_5)^a_{\ph{a}b}\epsilon^b \ = \ \pm \epsilon^a\, .
\eeq
This condition is compatible with the gravitino and dilatino equations \eqref{BulkGravitinoAHatzero} and \eqref{BulkDilatinoAHatzero}, since $\Gamma_5$ commutes with the self-dual 't Hooft symbols. Projecting onto the subspaces with positive or negative ``internal chirality'' in (\ref{eq:InternalChiralityEq}) further reduces the bulk spinor to
\begin{align}
	\epsilon^a \ =& \ \left( \zeta \, , -\zeta^c \, , \pm \ii \zeta \, , \mp \ii \zeta^c \right)^{\mathrm{T}} \, .
\end{align}

Using the single Dirac spinor $\zeta$, we may define the following (local) differential forms
\begin{gather}
 S \ \equiv  \ \overline{\zeta}\zeta\, , \qquad  \qquad \qquad \qquad  P \ \equiv \ \overline{\zeta}\Gamma_5\zeta\, , \nn \\
K \ \equiv \ \frac{1}{S}\overline{\zeta}\Gamma_{(1)}\zeta \, , \qquad V^1\mp \ii V^3 \ \equiv \frac{\ii}{S} \overline{\zeta^c}\Gamma_{(1)}\Gamma_5\zeta \, , \qquad   V^2 \ \equiv  \ \frac{\ii}{S}\overline{\zeta}\Gamma_{(1)}\Gamma_5\zeta \, ,
\end{gather}
where a bar denotes Hermitian conjugation. Globally, the full bulk spinor is a section of $Spin(\B_4)\otimes E$, where $E$ is a real rank 4 vector bundle associated to the principal $SU(2)_R$ bundle. By considering the change between local trivializations of the spinor under the $SU(2)_R \subset Spin(4)$, one can check that $S$ and $P$ are global smooth functions. Moreover, $K$ is a global one-form on $\B_4\setminus\{S=0\}$, whilst $(V^1,V^2,V^3)$ are sections of $\Omega^1(\B_4\setminus\{S=0\})\otimes V$, where $V$ is the rank 3 vector bundle associated to the $SO(3)_R = SU(2)_R/\Z_2$.

In order to have a globally well-defined bulk spinor $\epsilon^a$, we have to lift the $SO(3)_R$ bundle acting on $V$ to an $SU(2)_R$ bundle acting on $E$. Moreover, we should define the spinor in the first place, thus lifting the orthonormal frame bundle of the tangent bundle to a $Spin(4)$ frame bundle. In both cases, the obstruction to the lifting is the second Stiefel--Whitney class of the real vector bundles, that is, $w_2(V), w_2(\B_4)\in H^2(\B_4,\Z_2)$. However, because the full bulk spinor is a section of $Spin(\B_4)\otimes E$, we only  need
\beq
w_2(V) \ =\  w_2(\B_4) \, ,
\eeq
in order for the tensor product of the ``virtual'' bundles to be defined. We say that the bulk spinor is a $Spin_{SU(2)}$ spinor, as originally introduced in \cite{Back:1978zf}. This is a non-Abelian generalization of the perhaps more familiar $Spin^c$ spinors that are required, for instance, in Seiberg--Witten theory.

Geometrically, a single Dirac spinor in four dimensions defines a local identity structure on the four-manifold, or equivalently a local orthonormal frame. In order to construct it, we split the bulk spinor into its components with positive and negative chirality under $\Gamma_5$, $\zeta = \zeta_+ + \zeta_-$, and define
\beq
\eta_{\pm} \ \equiv \ \frac{\zeta_{\pm}}{\sqrt{S_{\pm}}}\, ,
\eeq
where $S_{\pm} \equiv \overline{\zeta_{\pm}}\zeta_{\pm}$. Then an orthonormal frame  can be defined by
\beq
\ii \Ex^2 - \Ex^4 \ \equiv \ \overline{\eta_-}\Gamma_{(1)} \eta_+ \, , \qquad \ii\Ex^1 - \Ex^3 \ \equiv \ \overline{\eta^c_-}\Gamma_{(1)}\eta_+ \, , 
\eeq
and we choose the orientation induced by the volume form $\Ex^{4123}$. We also define the function $\theta$ by
\beq
\cos^2\frac{\theta}{2} \ \equiv \ \frac{S_+}{S}\, , \qquad  \sin^2\frac{\theta}{2} \ \equiv \ \frac{S_-}{S}\, .
\eeq
We may then re-express the local differential forms above in terms of the frame as 
\begin{gather}
P \ = \ S \cos\theta\, , \qquad K \ = \ - \sin\theta\, \Ex^4\, , \qquad V^I \ = \ - \sin\theta \, \Ex^I\, , \quad I=1,2,3\, .
\end{gather}
This canonical frame degenerates at $\theta=0,\pi$, where the spinor has positive/negative chirality, and also when $S=0$, where the spinor vanishes. 
The subset of $\B_4$ with these points excluded will be denoted $\B_4^{(0)}$. From the global considerations above it then follows that 
 $\Ex^4$ is a global one-form on $\B_4^{(0)}$, and $\Ex^I$ are sections of $\Omega^1(\B_4^{(0)})\otimes V$. Therefore, the $\Ex^I$ rotate into each other in the fundamental representation of $SO(3)_R$ between local trivializations, and the orthonormal frame is not in general global.

Starting with the bulk Killing spinor equations \eqref{BulkGravitinoAHatzero} and \eqref{BulkDilatinoAHatzero}, we may find a set of Killing spinor equations for $\zeta$. Choosing negative internal chirality in \eqref{eq:InternalChiralityEq}, they read
\begin{align}
\nabla_{\mu}\zeta \ =& \ - \tfrac{\ii}{2\sqrt{2}}\cA^2_{\mu}\zeta - \tfrac{\ii}{2\sqrt{2}}(\cA^1_{\mu}+ \ii \cA^3_{\mu})\zeta^c + \tfrac{\ii}{8\sqrt{2}} X^{-1} \cF^2_{\nu\lambda}\Gamma^{\nu\lambda}\Gamma_{\mu}\zeta - \tfrac{\ii}{4}X^2\partial_{\mu}\j\, \Gamma_5\zeta \nn \\
& \ + \tfrac{\ii }{8\sqrt{2}}X^{-1}(\cF^1_{\nu\lambda} + \ii \cF^3_{\nu\lambda})\Gamma^{\nu\lambda}\Gamma_{\mu}\zeta^c
 - \tfrac{1}{4}(X+X^{-1})\Gamma_{\mu}\zeta - \tfrac{\ii}{4}\j X \Gamma_{\mu}\Gamma_5\zeta \, , \label{eq:SingleDiracGravitino}
\\[10pt]
0 \ =& \ \tfrac{1}{\sqrt{2}}X^{-1}\partial_{\nu}X\Gamma^{\nu}\zeta + \tfrac{\ii}{8} X^{-1}\cF^2_{\nu\lambda}\Gamma^{\nu\lambda}\zeta + \tfrac{\ii}{8}X^{-1}(\cF^1_{\nu\lambda} + \ii \cF^3_{\nu\lambda})\Gamma^{\nu\lambda}\zeta^c \nn \\
& \ - \tfrac{\ii}{2\sqrt{2}}X^2 \partial_{\nu}\j \Gamma^{\nu}\Gamma_5\zeta - \tfrac{1}{2\sqrt{2}}(X-X^{-1})\zeta - \tfrac{\ii}{2\sqrt{2}}\j X \Gamma_5\zeta\, . \label{eq:SingleDiracDilatino}
\end{align}

From these equations, one can use standard spinor bilinear manipulations to obtain differential conditions for the frame and the fields:
\begin{align}
\label{eq:defE4}
\dd(XS) \ =& \ S\sin\theta \, \Ex^4 \, , \\[10pt]
\label{eq:EtildeIFI}
\dd (XS\cos\theta) \ =& \ \tfrac{1}{\sqrt{2}} S \sin\theta  \, \Ex_I\hook \cF^I \, ,\\[10pt]
\label{eq:DEtildeI}
\begin{split}
- D(S\sin\theta \,\Ex^I) \ =& \ \tfrac{1}{\sqrt{2}} X^{-1}S(*\cF^I - \cos\theta \cF^I)\\
& \ +(X+X^{-1})S\left(\Ex^{I4} - \tfrac{1}{2}\cos\theta \, \epsilon^{IJK}\Ex^{JK}\right) \\
& \ +\ii\j XS \left( \cos\theta\, \Ex^{I4} - \tfrac{1}{2}\epsilon^{IJK}\Ex^{JK}\right) \, ,
\end{split} \\[10pt]
\label{eq:dPhione}
\dd\j \ =& \ \tfrac{\ii}{\sqrt{2}} X^{-3} \csc \theta \, {\Ex}_J\hook\left( \cF^J + \cos\theta *\cF^J\right) \nn \\
& \ + X^{-3}\csc \theta \left(\ii X(X-X^{-1})\cos\theta - \j X^2\right){\Ex}^4 \, ,\\[10pt]
\label{eq:dXone}
\dd X \ =& \ -\tfrac{1}{2\sqrt{2}} \csc \theta \, {\Ex}_J\hook\left( \cos\theta \, \cF^J + *\cF^J\right) \nn \\
& \ - \tfrac{1}{2} \csc \theta \left(X(X-X^{-1}) - \ii\j X^2\cos\theta\right){\Ex}^4 \, .
\end{align}
Here the covariant derivative acting on $\Ex^I$ is $D \Ex^I\equiv\diff \Ex^I -\tfrac{1}{\sqrt{2}}\epsilon^{IJK}\cA^J\wedge \Ex^K$.
We may in particular combine these equations to obtain an expression for $\j$:
\begin{align}\label{eq:PhiSUSY}
\j \ &= \ \ii X^{-2} \cos\theta + \alpha (XS)^{-1}\, ,
\end{align} 
where $\alpha\in\ii\R$, and we have used that everything in this last equation is globally defined to integrate, assuming that $\B_4$ is path-connected.

The system of equations (\ref{eq:defE4})--(\ref{eq:dXone}) is in fact necessary and sufficient to have a supersymmetric solution 
to the bulk equations of motion. There are several steps involved in showing this. Firstly, we note that for a Dirac spinor $\zeta$ 
the set $\{\zeta,\zeta^c,\Gamma_\mu\zeta\,\Gamma_\mu\zeta^c\}$ spans the spinor space. Thus contracting the dilatino equation 
(\ref{eq:SingleDiracDilatino}) with the Hermitian conjugate of each element of this set gives a collection of equations which are 
equivalent to the dilatino equation. 
In turn, these equations can be shown to be equivalent to (\ref{eq:dPhione}) and (\ref{eq:dXone}). On the other hand, since we have a (local)
identity structure, the intrinsic torsion is determined by the exterior derivatives in (\ref{eq:defE4})--(\ref{eq:DEtildeI}). It follows that (\ref{eq:defE4})--(\ref{eq:dXone})  
are equivalent to the Killing spinor equations. One next considers  the truncated integrability conditions derived from \eqref{eq:IntegrabilityDilatino} and \eqref{eq:IntegrabilityGravitino}. From these it is  straightforward to show that the Killing spinor equations imply the equations of motion, while the Bianchi identity for $\mc{F}^I$ has to be imposed additionally. In particular the proof of this uses the fact that the bulk spinor $\zeta$ is Dirac. 
The upshot is that the complete system of equations to solve is given by the first order differential system (\ref{eq:defE4})--(\ref{eq:dXone}).

It is interesting, especially in light of the computation of the on-shell action in the next section, to consider the expansion of the bilinear equation near the boundary. Using the Fefferman--Graham coordinate $z$, the bulk spinor $\zeta$ has the expansion 
\beq\label{eq:ExpansionSpinor}
\zeta \ = \ z^{-1/2} \begin{pmatrix}
\chi \\ 0
\end{pmatrix} + z^{3/2}\begin{pmatrix}
\frac{1}{16}R\, \chi \\ \frac{1}{2}\j_2\, \chi
\end{pmatrix}  + z^{5/2} \begin{pmatrix}
-\tfrac{1}{12}\partial_{\overline{i}}\j_2 \, \sigma^{\overline{i}}\chi \\
\tfrac{1}{16}\partial_{\overline{i}}R \, \sigma^{\overline{i}}\chi
\end{pmatrix} + o(z^{5/2}) \, ,
\eeq
where $\chi$ is a constant 2-component spinor given by
\beq
\chi \ = \ \begin{pmatrix}
c \\ -\ii c
\end{pmatrix}\, ,
\eeq
with $c\in\R$ (compare with \eqref{twistsoln} with $c = -\overline{w}$). Without loss of generality, we may set $c=1$ in the following, and the norm of the spinor takes the form
\beq\label{eq:ExpansionS}
S \ = \ \frac{2}{z} + \frac{z}{4}R + o(z^2) \, .
\eeq
We also find
\beq\label{eq:ExpansionFields}
\begin{split}
X \ &= \ 1 - \frac{z^2}{4}R + o(z^3) \, , \\
\j \ &= \ \frac{\ii}{\sqrt{2}}z^2 (a^I_1)_I + o(z^3) \, , \\
t^{(2)} \ &= \ - \frac{1}{4}R \, , \qquad t^{(3)} \ = \ 0 \, .
\end{split}
\eeq
The vanishing of $\j_1$ allows us to fix the constant $\alpha$ in \eqref{eq:PhiSUSY}: expanding the latter equation leads to $\j_1 = \alpha/2$, so under the assumption of the topological twist, $\alpha=0$. In a neighbourhood of the conformal boundary, the bulk frame has the form
\beq
\begin{split}
\Ex^I &= \frac{1}{z}\bdrye^I + \frac{z}{2}\left(\mexp^2\circ\bdrye^I\right) + o(z) \, ,  \\
\Ex^4 &= - \frac{\dd z}{z} - \frac{z^2}{8}\dd R + o(z^2) \, .
\end{split}
\eeq
Near the boundary, the leading order of the equations \eqref{eq:defE4}--\eqref{eq:dXone} is trivial apart from \eqref{eq:DEtildeI}, which corresponds to the condition that $\bdrye^I$ satisfy the first Cartan's structural equation
\beq\label{Cartan}
\dd\bdrye^I +\omega^{I}_{\ J}\wedge \bdrye^J \ = \ 0~.
\eeq
Here the spin connection $\omega^{I}_{\ J}$ arises from the topological twist boundary condition for the gauge field (\ref{twistcond}).
In some sense (\ref{Cartan}) is a redundant equation, simply stating that the frame defined by supersymmetry is compatible with the boundary metric. As in the AdS$_5$/CFT$_4$ example, the bulk differental equations are tautological on the boundary, where they simply define a (twisted) frame for the three-manifold $M_3$.  The analogous statement in AdS$_5$/CFT$_4$ is that the bulk differential system at the boundary defines a quaternionic K\"ahler structure on the supersymmetric background, which one can construct on any four-manifold \cite{BenettiGenolini:2017zmu}.

\subsection{On-shell action}\label{Sec3dOS}

Thanks to these results, we can now greatly simplify the expression for the on-shell action. We start with the expression \eqref{IEuclidOnShell} and set $\hat{\cF}^I=0$, obtaining
\beq
I_{\text{o-s}} \ = \ -\frac{1}{2\kappa_4^2}\int_{\B_4}\big[- (4 + X^2 + X^{-2}+\j^2 X^2) *1  - \tfrac{1}{2}X^{-2} \big( \cF^I\wedge *\cF^I + \ii \j X^2 \cF^I \wedge \cF^I \big)  \big] \, .
\eeq
Then, using \eqref{Xeom} and \eqref{AxionEOM}, we may exchange the gauge field contribution for an exact term
\beq\label{eq:IEintermediate}
I_{\text{o-s}} \ = \ -\frac{1}{2\kappa_4^2}\int_{Y_4}\big[- (4 + 2X^{-2}+ 2\j^2 X^2) *1  + \dd \big( 2X^{-1}*\dd X - \j X^4*\dd\j \big)  \big] \, .
\eeq
Notice that, using the equations for the orthonormal frame and \eqref{eq:PhiSUSY}, we can write
\beq
\dd \left(X^{-1}*K\right) \ = \ -\left( 2 + X^{-2} \sin^2\theta\right) *1 \, ,
\eeq
and this, using the expression \eqref{eq:PhiSUSY} for $\st$, is exactly (modulo a numerical factor) the potential term in the on-shell action \eqref{eq:IEintermediate}. Therefore, the on-shell action is exact
\beq\label{eq:ExactOnShell}
I_{\text{o-s}} \ = \ - \frac{1}{\kappa^2_4}\int_{\B_4}\dd \left( X^{-1}*K + X^{-1}*\dd X - \tfrac{1}{2}\j X^4* \dd\j \right)  \, .
\eeq
The global arguments discussed above imply that the four-form in the action
\beq
\Upsilon \ \equiv \ X^{-1}*K + X^{-1}*\dd X - \tfrac{1}{2}\j X^4* \dd\j \, ,
\eeq 
is globally well-defined on $\B_4^{(0)}$. In what follows we assume that the subset of $\B_4$ where the spinor becomes chiral or zero is 
measure zero. As in section \ref{SecHoloRenormalization}, we cut off the bulk $\B_4$ at some small radius $z=\delta>0$, so that 
$\partial \B_4 = M_\delta\equiv \{z=\delta\}\cong M_3$. 
Using Stokes' theorem, we may then write the on-shell action as integrals over the conformal boundary $M_3\cong M_{\delta}$, and over the boundaries $\mathrm{T}_{\epsilon}$ of the small tubular neighbourhoods of radius $\epsilon>0$ surrounding the subsets $\B_4\setminus\B_4^{(0)}$ where the frame degenerates. Let us consider first the contribution from the conformal boundary: using the expansion of the spinor \eqref{eq:ExpansionSpinor} and of the fields \eqref{eq:ExpansionFields}, it is easy to show that near the conformal boundary
\beq
\Upsilon \ = \ \left( \frac{1}{\delta^3} - \frac{3}{8\delta}R + o(1) \right)*_{\mexp^0}1 \, .
\eeq
To this we should add the contributions from the Gibbons--Hawking--York term \eqref{eq:GHY} and the counterterms \eqref{ctaction}, which in a neighbourhood of the boundary are 
\begin{align}
I_{\rm GHY} \ &= \ \frac{1}{\kappa^2_4}\int_{M_3}\left( - \frac{3}{\delta^3} + \frac{1}{8\delta}R + o(1) \right)*_{\mexp^0}1\, , \\[10pt]
I_{\text{ct}} \ &= \ \frac{1}{\kappa^2_4}\int_{M_3} \left(\frac{2}{\delta^3} + \frac{1}{4\delta}R + o(1) \right)*_{\mexp^0}1 \, .
\end{align}
Once we take into account the change in sign of the on-shell terms, due to the orientation of the bulk compared to the orientation of the boundary, the contribution to the renormalized action from the conformal boundary is zero in the limit $\delta\rightarrow 0$.

Therefore, the renormalized gravitational action only receives contributions from the subsets where the frame degenerates:
\beq
\Sren \ = \  \frac{1}{\kappa^2_4}\, \lim_{\epsilon\rightarrow 0} \int_{\text{T}_\epsilon}\Upsilon \, ,
\eeq
where the limit collapses the small neighbourhood around the degeneration locus. However, this gives zero for a smooth solution. That is, a supergravity solution with a smooth metric and smooth bosonic fields. Clearly the last two forms in $\Upsilon$, which only involve $X,\j$, are well-defined if the bosonic fields are smooth. In particular since $X=\ex^{\frac{1}{2}\phi}$, this means that 
$X>0$ (indeed, bounded below by a positive constant since $\B_4$ is compact). The last two terms in $\Upsilon$ therefore provide zero contribution when integrated over a subset of vanishing measure. The only non-trivial contribution could arise from $X^{-1}*K$.

Consider first the subset where the spinor is chiral but non-vanishing.  While changing from local $SU(2)_R$ gauge patches of definition for $\epsilon^a$, $\zeta$ is a linear combination of $\zeta$ and $\zeta^c$, but note that in four dimensions $\Gamma_5\zeta = \pm \zeta$ if and only if $\Gamma_5\zeta^c = \pm \zeta^c$. Therefore, spacetime chirality is a well-defined global concept for the $Spin_{SU(2)}$ spinor. If the spinor is chiral but non-vanishing, $S\neq 0$ and the bilinears $K$ and $V^I$ vanish, so $X^{-1}*K$ is zero there, and the integral is zero.

Secondly, consider the subset where the spinor is vanishing. One might worry that $K$ is not well-defined here, as $S=0$. However,
note that we may write
\beq
X^{-1}*K = - X^{-1} \sin\theta \, \Ex^4\hook \vol_{4}\, .
\eeq
Using (\ref{eq:defE4}) we then in turn have
\beq
X^{-1} \sin\theta \, \Ex^4 \ = \ \diff \log \rho~, \qquad \mbox{where} \quad \rho \ \equiv \ XS \, .
\eeq
We may thus use $\rho>0$ as a radial coordinate near to the where the spinor vanishes at $\rho=0$, and more precisely define 
$\text{T}_{\epsilon}=\{\rho=\epsilon>0\}$. It follows that $X^{-1}*K$ is 
the product of a bounded function $X^{-1}\sin\theta$ (as long as $X>0$ is smooth), and the volume form $\Ex^4\hook \vol_4$ induced on $\text{T}_\epsilon$ 
from the four-dimensional bulk metric. 
 The integral hence vanishes in the limit $\epsilon\rightarrow0$, where the volume of the tubular neighbourhood $\text{T}_\epsilon$ vanishes.

We conclude that the renormalized action for any \emph{smooth} supergravity solution is zero. In particular, 
we have made no assumptions at all here on the topology of $M_3$, or of its path-connected filling $Y_4$ with $\partial Y_4 = M_3$.

\section{Revisiting topological AdS$_5$/CFT$_4$}\label{SecRevisit}
Inspired by the evaluation of the on-shell action in the previous section, here we revisit the 5d/4d correspondence of \cite{BenettiGenolini:2017zmu}. After some brief background in section \ref{Sec5dBackground} recalling the work in \cite{BenettiGenolini:2017zmu}, we show in section \ref{Sec5dAction} that smooth supersymmetric five-dimensional bulk gravity fillings likewise have zero action.
Note that this section is entirely independent of the rest of the paper, despite sharing considerable overlap in notation. We trust this will not cause confusion.

\subsection{Background}\label{Sec5dBackground}

In \cite{BenettiGenolini:2017zmu} we defined a holographic dual to the Donaldson--Witten topological twist of $\mathcal{N}=2$ gauge theories on a Riemannian four-manifold $(M_4,g)$. The duals are described by a class of asymptotically locally hyperbolic solutions to (Euclidean) $\mathcal{N}=4^+$ gauged supergravity in five dimensions. Working in a truncation where a certain doublet of two-forms are set to zero, and with appropriate boundary conditions, we showed that the holographically renormalized on-shell action is independent of the boundary metric.

The action for the truncated Euclidean $\mathcal{N}=4^+$ gauged supergravity is
\begin{equation}\label{5dIEuclid}
\begin{split}
	I \ = \  - \frac{1}{2\kappa_5^2} \int_{\B_5} \ \Big[ &R \, {*1} - 3 X^{-2} \dd X \wedge *\dd X + 4  ( X^2 + 2 X^{-1} ) \, {*1} - \tfrac{1}{2} X^4 \, \cF \wedge *\cF  \\
	&- \tfrac{1}{4} X^{-2} \, \cF^I \wedge * \cF^I - \tfrac{\ii}{4} \cF^I \wedge \cF^I \wedge \cA \Big] \, .
	\end{split}
\end{equation}
Here $R=R(G)$ denotes the Ricci scalar of the five-dimensional metric $G_{\mu\nu}$,  $*$ is the Hodge duality operator acting on forms and $\cF = \dd \cA$, $\cF^I = \dd \cA^I - \tfrac{1}{2} \epsilon^I{}_{JK} \cA^J \wedge \cA^K$.
The equations of motion which follow from this action are:
\begin{align}
	\dd ( X^{-1}\, {*\, \dd X})   \  = & \ \tfrac{1}{3} X^4 \, \cF \wedge *\cF - \tfrac{1}{12} X^{-2} \, \cF^I \wedge *\cF^I - \tfrac{4}{3}  (X^2 - X^{-1})\, *1 \, , \label{5dXeom} \\[10pt]
	\dd ( X^{-2} * \cF^I ) \ =& \  \   \epsilon^{IJK} X^{-2} * \cF^J \wedge \cA^K - \ii \cF^I \wedge \cF \, , \label{5dAIeom} \\[10pt]
	\dd ( X^4 * \cF ) \ =& \ - \tfrac{\ii}{4} \cF^I \wedge \cF^I \, , \label{5dAeom} \\[10pt]	
	R_{\mu\nu}  \ = & \  3  X^{-2} \partial_\mu X \partial_\nu X - \tfrac{4}{3} ( X^2 + 2 X^{-1} ) G_{\mu\nu}+ \tfrac{1}{2} X^4 \big( \cF_\mu{}^\rho \cF_{\nu\rho} - \tfrac{1}{6} G_{\mu\nu} \cF^2 \big) \ \ \ \ \ \ \nn \\
	& + \tfrac{1}{4} X^{-2} \big( \cF^I_\mu{}^\rho \cF^I_{\nu\rho} - \tfrac{1}{6} G_{\mu\nu} ( \cF^I )^2 \big) \label{5dgeom}\, ,
\end{align}
with $\cF^2\equiv \cF_{\mu\nu}\cF^{\mu\nu}$ and $( \cF^I )^2\equiv \sum_{I=1}^3\cF^I_{\mu\nu}\cF^{I\mu\nu}$. 
Note that the one-form $\cA$ here plays a similar role to the axion $\j$ in the bulk four-dimensional supergravity of sections \ref{SecSUGRA}--\ref{SecGeometric}. In particular in \cite{BenettiGenolini:2017zmu} the field $\cA$ was likewise taken to be purely imaginary, with all other bosonic 
fields real.

A Fefferman--Graham expansion of the bosonic fields, together with imposing 
boundary conditions appropriate to the 4d $\mathcal{N}=2$ Donaldson--Witten topological twist, leads to the inverse radial coordinate expansions\footnote{For details of the metric expansion we refer the reader to \cite{BenettiGenolini:2017zmu}.}
\begin{align}
	X \ =& \ 1 -\tfrac{1}{12} z^2\log z  \, R + z^2 \Xs + \tfrac{1}{48}  z^4\log z\,  \nabla^2 R \nn \\
	&\ + z^4 \Big( - \tfrac{1}{4} \nabla^2 X_2 - \tfrac{1}{48} \nabla^2 R + \tfrac{1}{288} R^2 - \tfrac{1}{48} R_{ij} R^{ij} - \tfrac{1}{192} ( \mc{E} + \mc{P} ) \Big) + o(z^4) \, , \nn\\[10pt]
	\cA^I \ =&  \ \tfrac{1}{2}\omega_i{}^{jk}\Jb^I_{jk}\, \diff x^i - \tfrac{1}{4}z^2\log z \, \Jb^I_{mn}\nabla_j R^{mnj}_{\ \ \ \ \,  i}\, \diff x^i + z^2 \aIfive+o(z^2) \, ,  \nn\\[10pt]
	\cA\ =& \ z^2\, \mafive+ o(z^2)\, .
\end{align}
Here $R, R_{ij}$ and $R_{mnij}$ are respectively the boundary Ricci scalar, Ricci and Riemann tensor and $\mc{E}, \mc{P}$ are the Euler and Pontryagin densities constructed from these curvature tensors. The boundary spin connection is $\omega_i{}^{jk}$ and $\Jb^I$ are a triplet of boundary self-dual two-forms. 
The purely imaginary global one-form $\mafive$, along with $\Xs$ and $\aIfive$, are not determined by Fefferman--Graham expansion of the equations of motion.

As explained in section 5.1 of \cite{BenettiGenolini:2017zmu}, starting from a quadruplet of bulk spinors we may consistently truncate down to a single spinor $\zeta$ defined on $\B_5$ which satisfies
\begin{align}
\nabla_{\mu}\zeta \ =& \  - \tfrac{\ii}{2}\cA_{\mu}\zeta + \tfrac{\ii}{{2}} \left(\cA^1_{\mu} - \ii \cA^2_{\mu}\right)\zeta^c - \tfrac{\ii}{{2}} \cA^3_{\mu}\zeta + \tfrac{1}{3} \big( X + \tfrac{1}{2} X^{-2} \big) \gamma_\mu\zeta \nn\\
& \ + \tfrac{\ii}{24}  X^{-1} (\cF_{\nu\rho}^1-\ii \cF^2_{\nu\rho}) ( \gamma_\mu{}^{\nu\rho} - 4 \delta_\mu^\nu \gamma^\rho )\zeta^c - \tfrac{\ii}{24}  \big( X^{-1}\cF^3_{\nu\rho} + X^2 \mathcal{F}_{\nu\rho} \big) ( \gamma_\mu{}^{\nu\rho} - 4 \delta_\mu^\nu \gamma^\rho )\zeta
\, ,\nn\\[10pt]
0 \ =& \ \tfrac{{3\ii}}{2} \,  X^{-1} \partial_\mu X \gamma^\mu \zeta +{\ii}\big( X - X^{-2} \big)\zeta - \tfrac{1}{8} X^{-1} (\cF_{\mu\nu}^1-\ii \cF^2_{\mu\nu}) \gamma^{\mu\nu}\zeta^c \nn \\
& \ + \tfrac{1}{8} ( X^{-1}\cF^3_{\mu\nu} - {2} X^2 \mathcal{F}_{\mu\nu} )\gamma^{\mu\nu}\zeta\, . \label{5dsusy}
\end{align}
The five-dimensional charge conjugate condition is defined by $\zeta^c\equiv\mathscr{C}\zeta^*$.
From this single spinor we may define the following (local) differential forms
\begin{equation}
\label{eq:5dBilinears}
\begin{split}
S & \ \equiv \ \bar\zeta\zeta~, \qquad \qquad \quad \ \  \qquad \BK \ \equiv\  \frac{1}{S}\bar\zeta\gamma_{(1)}\zeta~, \\
\BJ^3 &\ \equiv \ \frac{\ii}{S}\bar\zeta\gamma_{(2)}\zeta, \ \quad \quad \BJ^2 + \ii \BJ^1 \ \equiv  \ \frac{1}{S}\bar\zeta^c\gamma_{(2)}\zeta \, ,
\end{split}
\end{equation}
which together constitute a twisted $Sp(1)$ structure. Using the supersymmetry equations \eqref{5dsusy} we derived the following system of differential equations for these bilinear forms\footnote{We have corrected a factor in \eqref{BilJeqns} compared to v1 of \cite{BenettiGenolini:2017zmu}.}
\begin{align}
X^{-2} \BK \ =& \ \diff \log (XS) +\ii \cA ~,\label{Kdiff} \\
	\diff (S \BJ^I) \ =& \ - \ii \cA\wedge S \BJ^I + (2X+X^{-2}) \BK \wedge S\BJ^I + \epsilon^{I}_{\ JK} \cA^J \wedge S \BJ^K \nn\\
	&\ + \tfrac{1}{2} X^{-1}S\, (*\cF^I +  \BK\wedge \cF^I) \, . \label{BilJeqns}
\end{align}
Here the Hodge dual is constructed from the volume form $\vol_5=-\frac{1}{6}\mathcal{K}\wedge \BJ^I \wedge \BJ^I$. 

\subsection{On-shell action}\label{Sec5dAction}

In \cite{BenettiGenolini:2017zmu} we showed that the on-shell action could be rewritten using the Einstein equation as
\beq
\begin{split}
	I_{\text{o-s}} \ = \ \frac{1}{2\kappa_5^2} \int_{\B_5} \ & \big[ \tfrac{8}{3} ( X^2 + 2 X^{-1} ) \, {*1} +\tfrac{1}{3} X^4 \cF \wedge * \cF + \tfrac{1}{6} X^{-2} \cF^I \wedge * \cF^I +\tfrac{\ii}{4} \cF^I \wedge \cF^I \wedge \cA \big] \, . \label{5dIEuclidOnShell1}
	\end{split}
\eeq
However, by additionally using the scalar field equation \eqref{5dXeom} twice and \eqref{5dAeom} to rewrite the Chern--Simons term we arrive at the following simpler expression
\begin{align}
	I_{\text{o-s}} \ = \ \frac{1}{2\kappa_5^2} \int_{\B_5} \ & \big[ 8 X^{-1} *1 - \dd \big( 2 X^{-1} * \dd X - X^4 \cA \wedge * \cF \big) \big] \, . \label{5dIEuclidOnShell2}
\end{align}
Now with some simple manipulation of the differential system \eqref{Kdiff}--\eqref{BilJeqns} we can show that
\begin{align}
	\tfrac{1}{3} \dd ( X^{-2} \BJ^I \wedge \BJ^I ) \ =& \ - 8 X^{-1} *1 \, ,
\end{align}
and immediately conclude that the on-shell action is (locally) exact;
\begin{align}
	I_{\text{o-s}} \ = \ - \frac{1}{2\kappa_5^2} \int_{\B_5} \ & \dd \big( \tfrac{1}{3} X^{-2} \BJ^I \wedge \BJ^I + 2 X^{-1} * \dd X - X^4 \cA \wedge * \cF \big) \, . \label{5dIEuclidOnShell3}
\end{align}

In addition to $\cA$ being a global one-form, with $\cF$ a global two-form, we assume that $X>0$ is a smooth global function on $\B_5$. Further, note that $\BJ^I \wedge \BJ^I \propto * \BK$ and $\BK$ is fixed by \eqref{Kdiff} in terms of $X$, $\cA$ and $S$. Hence $\BK$ is globally defined as long as the spinor norm $S = \bar\zeta\zeta \neq 0$.  We should hence more precisely define 
$\B_5^{(0)}\equiv \B_5\setminus \{S=0\}$, so that $(S,\BK,\BJ^I)$ are well-defined 
on $\B_5^{(0)}$ and the gravity solution is smooth. In summary, the on-shell action is globally exact and we may use Stokes' theorem to conclude %
\begin{align}
	I_{\text{o-s}} \ = \ - \frac{1}{2\kappa_5^2} \int_{\partial \B_5^{(0)}} \ & \left[\tfrac{1}{3} X^{-2} \BJ^I \wedge \BJ^I + 2 X^{-1} * \dd X - X^4 \cA \wedge * \cF \right] \, . \label{5dIEuclidOnShell4}
\end{align}
Here there are two types of boundary in $\partial \B_5^{(0)}$: firstly $\partial \B_5\cong M_4$ is the UV conformal boundary, and as in the previous
section there is also the boundary of a tubular neighbourhood $\text{T}_\epsilon$ around the locus where $\zeta=0$.

The above on-shell action must be supplemented by the standard Gibbons--Hawking--York term at the UV boundary, $I_{\text{GHY}}$ as in section \ref{SecHoloRenormalization}. In addition, the divergences may be cancelled by adding local boundary counterterms. The divergences identified by expanding \eqref{5dIEuclidOnShell1} and $I_{\text{GHY}}$ are cancelled by adding 
\beq\label{5dctaction}
\begin{split}
	I_{\text{ct}} \ = \ 
	\frac{1}{\kappa_5^2} \int_{\partial \B_5}  &\dd^4 x \, \sqrt{\det h}\, \Big\{ {3} +\tfrac{1}{4} R ( h ) +{3} ( X - 1 )^2  \\
	&+\log \cutoff \, \Big[- \tfrac{1}{8} \Big( R_{ij} ( h ) R^{ij} ( h ) - \tfrac{1}{3} R(h)^2 \Big)  +\tfrac{3}{2} ( \log \cutoff )^{-2} ( X - 1 )^2 \\
	&\qquad \qquad +\tfrac{1}{8} \cF^2_h + \tfrac{1}{16} ( \cF^I )^2_h \Big]\Big\} \, .
\end{split}
\eeq
Here the integral is over the UV boundary $\partial \B_5\cong M_4$ of $\B_5$.
As the on-shell actions given by \eqref{5dIEuclidOnShell1} and \eqref{5dIEuclidOnShell4} are equivalent, $I_{\text{ct}}$ must also cancel divergences arising from the latter when supplemented by the common Gibbons--Hawking--York term. The total renormalized action is then
\beq
\Sren_{\text{ren}} \ = \ \lim_{\cutoff\rightarrow 0} \, \left(I_{\text{o-s}} + I_{\text{GHY}} + I_{\text{ct}}\right) \, ,\label{Actionfinal5d}
\eeq
where $\delta$ is a cut--off for the radial coordinate $z$.

In order to calculate the UV contribution to $\Sren_{\text{ren}}$ of the term $\tfrac{1}{3} X^{-2} \BJ^I \wedge \BJ^I$ in $I_{\text{o-s}}$ we require the Fefferman--Graham-like expansion of the spinor $\zeta$ to one more order in $z$ than given in \cite{BenettiGenolini:2017zmu}. Continuing the line of reasoning there (and in section \ref{SecSUSYexpand})  we eventually compute\footnote{The $z^{5/2}$ term has been corrected compared to v1 of \cite{BenettiGenolini:2017zmu}.}
\begin{align}
	\zeta  \ =& \ z^{-1/2} \chi +z^{3/2} \left(\tfrac{1}{48} R \right)\chi +z^{5/2}\left( - \tfrac{1}{24} \diff R \, \log z + \tfrac{\ii}{2}\mafive + \tfrac{1}{2}\diff \Xs + \tfrac{1}{48}\diff R\right)_i \gamma^i \chi \nn \\
	&\ + z^{7/2} \Big[ - \tfrac{1}{1152} R^2 \, \log^2 z + \tfrac{1}{48} \left( R \Xs + \tfrac{1}{16} R^2 - \tfrac{1}{4} \nabla^2 R \right) \log z \nn \\
	&\ \qquad \quad \ - \tfrac{1}{8} \left( \Xs^2 + \tfrac{1}{8} R \Xs + \tfrac{1}{128} R^2 - \tfrac{1}{96} \nabla^2 R - \tfrac{1}{24} R_{ij} R^{ij} - \tfrac{\ii}{12} (\dd \mafive)_{ij} \gamma^{ij} \right) \Big] \chi \nn \\
	&\ + z^{7/2} \Big[ \tfrac{\ii}{96} \left( \mathcal{D} \aIfive \right)_{ij} \gamma^{ij} ( \sigma_I \epsilon^{-1} )_1 \Big] + o(z^4) \, .
\end{align}
The last line, seemingly, cannot be written in terms of the lowest order constant spinor $\chi$, whose norm is one, however it will not play a part. From this expansion and the definition of the bilinears in \eqref{eq:5dBilinears} we determine
\begin{align}
	\left.\BJ^I\wedge\BJ^I\right|_{z=\delta} \ =& \ \Big[ \tfrac{6}{\cutoff^4} - \tfrac{1}{2\cutoff^2} R + \tfrac{1}{8}\left( \tfrac{1}{3} R^2 - R_{ij}R^{ij}\right) - \tfrac{1}{24} R^2 \, \log^2\cutoff + R \Xs \, \log\cutoff \nn \\
	& \ \  + \tfrac{1}{128}\left(-384 \Xs^2 + \mc{E} + \mc{P} \right) \Big] \vol_4 + o(\cutoff^{1/2}) \, .
\end{align}
Here we have restricted the two-forms to the boundary at constant $z=\cutoff$. On forming the exterior product there are several simplifications, in particular the anti-symmetric indices of $\dd \mafive$ and $\mathcal{D} \aIfive$ are traced over and do not contribute. This can also be shown by expanding the equation $\BK\wedge\BJ^I\wedge\BJ^I = \vol_5$. 

We are finally in a position to evaluate the UV contribution to the renormalized on-shell action \eqref{Actionfinal5d}. We find
\begin{align}
	\Sren_{\text{ren}}^{\text{UV}} \ = \ \lim_{\cutoff\rightarrow 0} \frac{1}{\kappa_5^2} \int_{\partial \B_5} \left[ \log \delta \Big( \tfrac{1}{32} ( \mc{E} + \mc{P} ) *_4 1 + \tfrac{1}{24} \dd *_4 \dd R \Big) - \tfrac{1}{48} \dd *_4 \dd ( R + 24 X_2 ) \right] \, . \label{SrenUV}
\end{align} 
At first sight the $\log \cutoff$ term is problematic as it diverges. However, the topological condition $\int_{\partial \B_5} ( \mc{E} + \mc{P} ) *_4 1 = 0$ is required in order for $\cA$ to be a global one-form, or equivalently to have a non-zero partition function for the boundary TQFT \cite{BenettiGenolini:2017zmu}. Moreover, the Ricci scalar is a globally defined function on $\partial \B_5$, and consequently for boundaryless four-manifolds,
{\it i.e.}\ $\partial ( \partial \B_5 ) =0$, the second term vanishes on using Stokes' theorem. The same argument applies to the finite piece of $\Sren_{\text{ren}}^{\text{UV}}$ as the bulk scalar $X$, and hence $X_2$, is a global smooth function. It follows that the UV contribution to the renormalized action is zero for smooth fillings. 

As in the previous section, that now leaves us with the contribution from the small tubular neighbourhood $\text{T}_\epsilon$:
\beq
\Sren_{\text{ren}} \ =  \ \frac{1}{\kappa_5^2}\, \lim_{\epsilon\rightarrow 0} \int_{\text{T}_\epsilon} \big[-\tfrac{1}{6}X^{-2}\wedge \BJ^I\wedge\BJ^I-X^{-1}*\diff X + \tfrac{1}{2}X^4 \cA\wedge *\cF \big]~.
\eeq
The contributions from the second and third forms are zero for smooth solutions, again since $X>0$ is smooth, and $\cA$ is assumed to be a 
global smooth one-form on $\B_5$. Thus the integrals tend to zero as the volume enclosed by $\text{T}_\epsilon$ tends to zero. 
On the other hand, the first term may be written as
\beq
-\tfrac{1}{6}X^{-2}\BJ^I\wedge \BJ^I \ = \ X^{-2}*\mathcal{K}~.
\eeq
That this also contributes zero may now be argued in exactly the same way as at the end of section \ref{Sec3dOS}, using 
equation (\ref{Kdiff}).

We conclude that the renormalized action for any \emph{smooth} supergravity solution is zero. 
In particular, since $\B_5$ is assumed to have boundary $\partial \B_5=M_4$, together 
with the topological constraint mentioned after equation (\ref{SrenUV}) one necessarily has 
Euler number and signature of $M_4$ equal to zero: $\chi(M_4)=0=\sigma(M_4)$. 
Apart from this, no other topological assumption is made about $M_4$ or its filling in the above computation.


\section{Discussion}\label{SecDiscussion}

In this paper we have defined and studied a holographic dual to the topological twist of $\mathcal{N}=4$ gauge theories on Riemannian three-manifolds and verified that the renormalized gravitational free energy is independent of the boundary three-metric, thus providing an additional construction of topological AdS/CFT beyond \cite{BenettiGenolini:2017zmu}. We have also reformulated the bulk supersymmetry equations in terms of a twisted identity structure, and used this structure to prove that the gravitational free energy of all smooth bulk fillings, irrespective of their topology, is zero. 
Let us again emphasize that the latter result does not make the former result of section \ref{SecVary} redundant: the computation of the variation of the 
gravitational free energy holds for smooth solutions, but {\it a priori} it is more general. Metric-independence will still hold for singular 
solutions, provided the additional surface terms around the singularities are zero. 
In fact if one allows singular saddle point solutions at all, 
this should be a clear constraint. In addition we have revisited the AdS$_5$/CFT$_4$ correspondence and similarly showed that smooth fillings there also have zero gravitational free energy. 
The results presented here and in \cite{BenettiGenolini:2017zmu} raise a number of interesting questions and directions for future research.

In general the classical supergravity limit of the AdS/CFT correspondence identifies 
\beq
-\log Z_{\text{QFT}} \ = \ \Sren_{\text{ren}}~.\label{Sholo}
\eeq
Here on the right hand side we have the least action solution to the given filling problem in the bulk supergravity, while the left hand side is understood to be the 
leading term in the corresponding strong coupling (typically large rank $N$) limit of the QFT partition function. 
For example, uplifting the four-dimensional $\mathcal{N}=4$ gauged supergravity solutions to M-theory on $S^7/\Z_k$ leads 
to the effective four-dimensional Newton constant in (\ref{4dNewton}), which scales as $N^{3/2}$. The latter multiplies the 
holographically renormalized on-shell action $\Sren_{\text{ren}}$ on the right hand side of (\ref{Sholo}). On the other hand, 
in this paper we have shown that this gravitational free energy is always zero, for any smooth supergravity filling 
of any conformal boundary three-manifold $M_3$. We have already noted that every oriented three-manifold is spin, 
but another important topological fact is that every such three-manifold bounds a smooth four-manifold (which may be taken to be spin). 
There is thus no topological obstruction to finding such a bulk filling of $M_3$. Of course, an important assumption 
here is that there exist smooth fillings that solve the supergravity equations, with prescribed conformal boundary $(M_3,g)$. 
We have recast the supergravity equations as the first order differential system (\ref{eq:defE4})--(\ref{eq:dXone}), and thus existence and uniqueness
theorems for solutions to these equations will play an important role. Given that such solutions are supersymmetric and are dual to  a topologically twisted theory, one naturally expects better behaviour than the non-supersymmetric Einstein filling problem, typically studied by mathematicians. 
In any case, assuming that such smooth fillings are the dominant saddle points in (\ref{Sholo}), the results of this paper 
imply that the large $N$ limit of the topologically twisted ABJM partition function is $o(N^{3/2})$, for any three-manifold $M_3$. 
This should be contrasted with the non-twisted partition function on (for example) $S^3$, where both sides of (\ref{Sholo}) 
agree and equal $\frac{\pi \sqrt{2k}}{3}N^{3/2}$ in the large $N$ limit \cite{Drukker:2010nc}. It thus remains an interesting open 
problem to compute the large $N$ limit of the topologically twisted ABJM theory, on a three-manifold $M_3$, and compare with 
our holographic result. Moreover, if the leading classical saddle point indeed contributes zero, the next obvious step is 
to try to compute the subleading term, as a correction to the supergravity limit. Since by construction everything is a topological
invariant, this may well be possible.

Similar remarks apply to the Donaldson--Witten twist studied holographically in \cite{BenettiGenolini:2017zmu}. 
Here the bulk five-dimensional $\mathcal{N}=4^+$ gauged supergravity solutions uplift on $S^5$ to solutions 
of type IIB supergravity, where now the five-dimensional Newton constant is given by $\frac{1}{\kappa^2_5}=\frac{N^2}{4\pi^2}$.\footnote{One 
may also uplift to solutions of M-theory, which are dual to $\mathcal{N}=2$ theories of class $\mathcal{S}$ with $N^3$ scaling, but we won't discuss this further here.}
The resulting solutions are holographically dual to the Donaldson--Witten twist of $\mathcal{N}=4$ SYM on the conformal boundary 
four-manifold $M_4$. Similar remarks apply to those made in the paragraph above, although there is an important 
difference: the partition function is only non-zero when $2\chi(M_4)+3\sigma(M_4)=0$, and moreover $M_4$ bounds 
a smooth five-manifold if and only if $\sigma(M_4)=0$. The fact that the gravitational free energy is zero for smooth fillings, 
as shown in section \ref{SecRevisit}, is therefore only directly applicable when $\chi(M_4)=0=\sigma(M_4)$. In this case, 
the topologically twisted partition function of $\mathcal{N}=4$ SYM should be $o(N^2)$, assuming the dominant saddle point
solution is indeed smooth. 

On the other hand, the Donaldson--Witten twisted partition function has been computed, for general rank gauge group 
$\mathscr{G}=SU(N)$, on $M_4=\mathrm{K3}$ in  \cite{Vafa:1994tf, Tanaka:2017bcw}. This follows from the fact that
on the hyperK\"ahler K3 manifold the Donaldson--Witten and Vafa--Witten twists are equivalent (and in fact equivalent to the untwisted 
theory).  However, $|\sigma(\mathrm{K3})| = 16$ and a smooth filling by $\B_5$ does not exist in this case, so there is no 
obvious classical gravity solution to compare to. Nevertheless, the partition function  is (for $N$ prime)
\cite{Vafa:1994tf, Tanaka:2017bcw}
\beq
Z({\mathrm{K3}}) \ = \ \frac{1}{N^2}G(q^N) + \frac{1}{N}\sum_{I=1}^N G\left(\omega^I q^{1/N}\right)~,\label{ZK3}
\eeq
where $q=\exp(2\pi \ii \tau)$, with $\tau=\frac{\theta}{2\pi}+\frac{4\pi \ii}{g_{\mathrm{YM}}^2}$ the usual complexified gauge coupling, $\omega=\exp(2\pi \ii/N)$, and 
$G(q)=1/\eta^{24}(\tau)$, with $\eta$ the Dedekind eta-function. Taking the 't Hooft coupling $\lambda=g^2_{\mathrm{YM}}N$ fixed and 
large, the $N\rightarrow \infty$ limit is dominated by the first term in (\ref{ZK3}), resulting in the leading order behaviour
\beq
\log Z(\mathrm{K3}) \ \sim \  \frac{8\pi^2N^2}{\lambda}~.\label{ZK3largeN}
\eeq
As mentioned above, in general the classical gravitational free energy is order $N^2$, which for smooth fillings of $M_4$ we have 
shown is multiplied by zero for the holographic Donaldson--Witten twist. However, there is no such smooth filling of $M_4=$ K3, so it is not clear what the dual classical solution should be. 
Perhaps one should allow for certain singular $\B_5$, and/or fill the boundary $S^5\times \mathrm{K3}$ with 
a topology that is not simply an $S^5$ bundle over $\B_5$. These would lie outside the class of smooth solutions 
to the consistently truncated five-dimensional $\mathcal{N}=4^+$ gauged supergravity we have studied. 
That said, a perhaps naive interpretation of (\ref{ZK3largeN}) is that the leading classical $O(N^2)$ term is 
indeed zero, with the $N^2/\lambda$ term being a subleading string correction to this. 
This particular example clearly deserves much further study.

More generally, there are a wide  variety of possible topologically twisted theories in diverse spacetime dimensions. One could ask if zero action/gravitational free energy for smooth supergravity solutions dual to TQFTs is a general property. Perhaps this is specific to cases in which the 
preserved supercharge $Q$ in the TQFT satisfies $Q^2=0$, which is  generally not the case. The apparent simplicity of our results 
suggests there should be a more elegant way to set up the holographic problem. Recall that in field theory, invariance of the TQFT partition function with respect to metric deformations crucially relies on the stress-energy tensor being $Q$-exact. We have shown the corresponding result 
holographically, but in a less direct manner. It is natural to conjecture that a topological sector of gauged supergravities, 
in this holographic setting, may be similarly described using a boundary BRST symmetry \cite{Witten:1988xi, Imbimbo:2018duh, Rosa:2018xpt, Bae:2015eoa, Val2018}.


\subsection*{Acknowledgments}

\noindent 
We would like to thank the organizers and participants of Eurostrings 2018 for many interesting comments and discussions, with particular thanks to 
Fernando Alday, Nikolay Bobev, Sergei Gukov, Elias Kiritsis, Greg Moore and Sameer Murthy.  We thank Val Reys for providing us with a draft of \cite{Val2018}. 
P.~R. was supported in part by an INFN Fellowship.



\begin{thebibliography}{}

\bibitem{Maldacena:1997re} 
  J.~M.~Maldacena,
  ``The Large N limit of superconformal field theories and supergravity,''
  Int.\ J.\ Theor.\ Phys.\  {\bf 38}, 1113 (1999)
  [Adv.\ Theor.\ Math.\ Phys.\  {\bf 2}, 231 (1998)]
  [hep-th/9711200].
  
  \bibitem{Witten:1998qj}
  E.~Witten,
  ``Anti-de Sitter space and holography,''
  Adv.\ Theor.\ Math.\ Phys.\  {\bf 2} (1998) 253
  [hep-th/9802150].
  
  \bibitem{Gubser:1998bc}
  S.~S.~Gubser, I.~R.~Klebanov and A.~M.~Polyakov,
  ``Gauge theory correlators from noncritical string theory,''
  Phys.\ Lett.\ B {\bf 428} (1998) 105
  [hep-th/9802109].

\bibitem{BenettiGenolini:2017zmu}
  P.~Benetti Genolini, P.~Richmond and J.~Sparks,
  ``Topological AdS/CFT,''
  JHEP {\bf 1712} (2017) 039
  [arXiv:1707.08575 [hep-th]].

\bibitem{Witten:1988ze} 
  E.~Witten,
  ``Topological Quantum Field Theory,''
  Commun.\ Math.\ Phys.\  {\bf 117}, 353 (1988).
  
\bibitem{Labastida:1998sk} 
  J.~M.~F.~Labastida and C.~Lozano,
  ``Duality in twisted N=4 supersymmetric gauge theories in four-dimensions,''
  Nucl.\ Phys.\ B {\bf 537}, 203 (1999)
  [hep-th/9806032].
  
\bibitem{Maloney:2007ud}
  A.~Maloney and E.~Witten,
  ``Quantum Gravity Partition Functions in Three Dimensions,''
  JHEP {\bf 1002} (2010) 029
  [arXiv:0712.0155 [hep-th]].  
  
\bibitem{Banerjee:2009af}
  N.~Banerjee, S.~Banerjee, R.~K.~Gupta, I.~Mandal and A.~Sen,
  ``Supersymmetry, Localization and Quantum Entropy Function,''
  JHEP {\bf 1002} (2010) 091
  [arXiv:0905.2686 [hep-th]].  
  
\bibitem{Alday:2012au} 
  L.~F.~Alday, M.~Fluder and J.~Sparks,
  ``The Large N limit of M2-branes on Lens spaces,''
  JHEP {\bf 1210}, 057 (2012)
  [arXiv:1204.1280 [hep-th]].
  
\bibitem{Witten:1992fb} 
  E.~Witten,
 ``Chern-Simons gauge theory as a string theory,''
  Prog.\ Math.\  {\bf 133}, 637 (1995)
  [hep-th/9207094].

\bibitem{Ooguri:2002gx} 
  H.~Ooguri and C.~Vafa,
  ``World sheet derivation of a large N duality,''
  Nucl.\ Phys.\ B {\bf 641}, 3 (2002)
  [hep-th/0205297].

\bibitem{Blau:1996bx}
  M.~Blau and G.~Thompson,
  ``Aspects of N(T) $\geq 2$ topological gauge theories and D-branes,''
  Nucl.\ Phys.\ B {\bf 492} (1997) 545
  [hep-th/9612143].
  
\bibitem{Das:1977pu}
  A.~Das, M.~Fischler and M.~Ro\v{c}ek,
  ``SuperHiggs Effect in a New Class of Scalar Models and a Model of Super QED,''
  Phys.\ Rev.\ D {\bf 16} (1977) 3427.
  
\bibitem{Cvetic:1999au} 
  M.~Cvetic, H.~Lu and C.~N.~Pope,
  ``Four-dimensional N=4, SO(4) gauged supergravity from D = 11,''
  Nucl.\ Phys.\ B {\bf 574}, 761 (2000)
  [hep-th/9910252].
  
\bibitem{Witten:1989sx}
  E.~Witten,
  ``Topology Changing Amplitudes in (2+1)-Dimensional Gravity,''
  Nucl.\ Phys.\ B {\bf 323} (1989) 113.

\bibitem{Birmingham:1989is}
  D.~Birmingham, M.~Blau and G.~Thompson,
  ``Geometry and Quantization of Topological Gauge Theories,''
  Int.\ J.\ Mod.\ Phys.\ A {\bf 5} (1990) 4721.

\bibitem{Blau:1991bn}
  M.~Blau and G.~Thompson,
  ``N=2 topological gauge theory, the Euler characteristic of moduli spaces, and the Casson invariant,''
  Commun.\ Math.\ Phys.\  {\bf 152} (1993) 41
  [hep-th/9112012].

\bibitem{Blau:2000iy}
  M.~Blau and G.~Thompson,
  ``On the relationship between the Rozansky-Witten and the three-dimensional Seiberg-Witten invariants,''
  Adv.\ Theor.\ Math.\ Phys.\  {\bf 5} (2002) 483
  [hep-th/0006244].  

\bibitem{Mikhaylov:2015nsa}
  V.~Mikhaylov,
  ``Analytic Torsion, 3d Mirror Symmetry And Supergroup Chern-Simons Theories,''
  arXiv:1505.03130 [hep-th].
  
\bibitem{Geyer:2001yc} 
  B.~Geyer and D.~Mulsch,
  ``N(T) = 4 equivariant extension of the 3-D topological model of Blau and Thompson,''
  Nucl.\ Phys.\ B {\bf 616}, 476 (2001)
  [hep-th/0108042].
  
\bibitem{Gaiotto:2008sd}
  D.~Gaiotto and E.~Witten,
  ``Janus Configurations, Chern-Simons Couplings, And The theta-Angle in N=4 Super Yang-Mills Theory,''
  JHEP {\bf 1006} (2010) 097
  [arXiv:0804.2907 [hep-th]].  
  
\bibitem{Kapustin:2009cd}
  A.~Kapustin and N.~Saulina,
  ``Chern-Simons-Rozansky-Witten topological field theory,''
  Nucl.\ Phys.\ B {\bf 823} (2009) 403
  [arXiv:0904.1447 [hep-th]].  
  
\bibitem{Koh:2009um}
  E.~Koh, S.~Lee and S.~Lee,
  ``Topological Chern-Simons Sigma Model,''
  JHEP {\bf 0909} (2009) 122
  [arXiv:0907.1641 [hep-th]].
 
\bibitem{Aharony:2008ug} 
  O.~Aharony, O.~Bergman, D.~L.~Jafferis and J.~Maldacena,
  ``N=6 superconformal Chern-Simons-matter theories, M2-branes and their gravity duals,''
  JHEP {\bf 0810}, 091 (2008)
  [arXiv:0806.1218 [hep-th]].
  
\bibitem{Lee:2008cr}
  K.~Lee, S.~Lee and J.~H.~Park,
  ``Topological Twisting of Multiple M2-brane Theory,''
  JHEP {\bf 0811} (2008) 014
  [arXiv:0809.2924 [hep-th]].  

\bibitem{Bagger:2006sk}
  J.~Bagger and N.~Lambert,
  ``Modeling Multiple M2's,''
  Phys.\ Rev.\ D {\bf 75} (2007) 045020
  [hep-th/0611108].
  
  \bibitem{Bagger:2007jr}
      J.~Bagger and N.~Lambert,
  ``Gauge symmetry and supersymmetry of multiple M2-branes,''
  Phys.\ Rev.\ D {\bf 77} (2008) 065008
  [arXiv:0711.0955 [hep-th]].
  
  \bibitem{Bagger:2007vi}
   J.~Bagger and N.~Lambert,
  ``Comments on multiple M2-branes,''
  JHEP {\bf 0802} (2008) 105
  [arXiv:0712.3738 [hep-th]].
  
  \bibitem{Gustavsson:2007vu}
   A.~Gustavsson,
  ``Algebraic structures on parallel M2-branes,''
  Nucl.\ Phys.\ B {\bf 811} (2009) 66
  [arXiv:0709.1260 [hep-th]].

\bibitem{Aharony:2008gk}
  O.~Aharony, O.~Bergman and D.~L.~Jafferis,
  ``Fractional M2-branes,''
  JHEP {\bf 0811} (2008) 043
  [arXiv:0807.4924 [hep-th]].

\bibitem{Lambert:2010ji}
  N.~Lambert and C.~Papageorgakis,
  ``Relating U(N)xU(N) to SU(N)xSU(N) Chern-Simons Membrane theories,''
  JHEP {\bf 1004} (2010) 104
  [arXiv:1001.4779 [hep-th]];

\bibitem{Bashkirov:2011pt}
  D.~Bashkirov and A.~Kapustin,
  ``Dualities between N = 8 superconformal field theories in three dimensions,''
  JHEP {\bf 1105} (2011) 074
  [arXiv:1103.3548 [hep-th]];

\bibitem{Agmon:2017lga}
  N.~B.~Agmon, S.~M.~Chester and S.~S.~Pufu,
  ``A New Duality Between $\mathcal{N}=8$ Superconformal Field Theories in Three Dimensions,''
  arXiv:1708.07861 [hep-th].
  
\bibitem{Mukhi:2008ux}
  S.~Mukhi and C.~Papageorgakis,
  ``M2 to D2,''
  JHEP {\bf 0805} (2008) 085
  [arXiv:0803.3218 [hep-th]].   
  
\bibitem{Festuccia:2011ws}
  G.~Festuccia and N.~Seiberg,
  ``Rigid Supersymmetric Theories in Curved Superspace,''
  JHEP {\bf 1106} (2011) 114
  [arXiv:1105.0689 [hep-th]].  
  
\bibitem{Klare:2012gn}
  C.~Klare, A.~Tomasiello and A.~Zaffaroni,
  ``Supersymmetry on Curved Spaces and Holography,''
  JHEP {\bf 1208} (2012) 061
  [arXiv:1205.1062 [hep-th]].  
  
\bibitem{Karlhede:1988ax}
  A.~Karlhede and M.~Ro\v{c}ek,
  ``Topological Quantum Field Theory and $N=2$ Conformal Supergravity,''
  Phys.\ Lett.\ B {\bf 212} (1988) 51.
  
\bibitem{Klare:2013dka}
  C.~Klare and A.~Zaffaroni,
  ``Extended Supersymmetry on Curved Spaces,''
  JHEP {\bf 1310} (2013) 218
  [arXiv:1308.1102 [hep-th]].   
  
\bibitem{Hristov:2013spa}
  K.~Hristov, A.~Tomasiello and A.~Zaffaroni,
  ``Supersymmetry on Three-dimensional Lorentzian Curved Spaces and Black Hole Holography,''
  JHEP {\bf 1305} (2013) 057
  [arXiv:1302.5228 [hep-th]].

\bibitem{Closset:2012ru}
  C.~Closset, T.~T.~Dumitrescu, G.~Festuccia and Z.~Komargodski,
  ``Supersymmetric Field Theories on Three-Manifolds,''
  JHEP {\bf 1305} (2013) 017
  [arXiv:1212.3388 [hep-th]].

\bibitem{Hosomichi:2008jb}
  K.~Hosomichi, K.~M.~Lee, S.~Lee, S.~Lee and J.~Park,
  ``N=5,6 Superconformal Chern-Simons Theories and M2-branes on Orbifolds,''
  JHEP {\bf 0809} (2008) 002
  [arXiv:0806.4977 [hep-th]].  

\bibitem{Fefferman:2007rka}
  C.~Fefferman and C.~Graham,
  ``The ambient metric,''
  arXiv:0710.0919 [math.DG].

\bibitem{Emparan:1999pm}
  R.~Emparan, C.~V.~Johnson and R.~C.~Myers,
  ``Surface terms as counterterms in the AdS / CFT correspondence,''
  Phys.\ Rev.\ D {\bf 60} (1999) 104001
  [hep-th/9903238];
   
\bibitem{Taylor:2000xw}
  M.~Taylor,
  ``More on counterterms in the gravitational action and anomalies,''
  hep-th/0002125;
 
\bibitem{deHaro:2000vlm}
  S.~de Haro, S.~N.~Solodukhin and K.~Skenderis,
  ``Holographic reconstruction of space-time and renormalization in the AdS / CFT correspondence,''
  Commun.\ Math.\ Phys.\  {\bf 217} (2001) 595
  [hep-th/0002230].
  
\bibitem{Genolini:2016sxe} 
  P.~Benetti Genolini, D.~Cassani, D.~Martelli and J.~Sparks,
  ``The holographic supersymmetric Casimir energy,''
  Phys.\ Rev.\ D {\bf 95}, no. 2, 021902 (2017)
  [arXiv:1606.02724 [hep-th]].
  
\bibitem{Genolini:2016ecx} 
  P.~Benetti Genolini, D.~Cassani, D.~Martelli and J.~Sparks,
  ``Holographic renormalization and supersymmetry,''
  JHEP {\bf 1702}, 132 (2017)
  [arXiv:1612.06761 [hep-th]].
  
\bibitem{Papadimitriou:2017kzw} 
  I.~Papadimitriou,
  ``Supercurrent anomalies in 4d SCFTs,''
  JHEP {\bf 1707}, 038 (2017)
  [arXiv:1703.04299 [hep-th]].

\bibitem{An:2017ihs} 
  O.~S.~An,
  ``Anomaly-corrected supersymmetry algebra and supersymmetric holographic renormalization,''
  JHEP {\bf 1712}, 107 (2017)
  [arXiv:1703.09607 [hep-th]].

\bibitem{Banerjee:2015uee} 
  N.~Banerjee, B.~de Wit and S.~Katmadas,
  ``The off-shell c-map,''
  JHEP {\bf 1601}, 156 (2016)
  [arXiv:1512.06686 [hep-th]].
 
\bibitem{Back:1978zf} 
  A.~Back, P.~G.~O.~Freund and M.~Forger,
  ``New Gravitational Instantons and Universal Spin Structures,''
  Phys.\ Lett.\  {\bf 77B}, 181 (1978).

\bibitem{Drukker:2010nc}
  N.~Drukker, M.~Marino and P.~Putrov,
  ``From weak to strong coupling in ABJM theory,''
  Commun.\ Math.\ Phys.\  {\bf 306} (2011) 511
  [arXiv:1007.3837 [hep-th]].

\bibitem{Vafa:1994tf}
  C.~Vafa and E.~Witten,
  ``A Strong coupling test of S duality,''
  Nucl.\ Phys.\ B {\bf 431} (1994) 3
  [hep-th/9408074].
  
\bibitem{Tanaka:2017bcw}
  Y.~Tanaka and R.~P.~Thomas,
  ``Vafa-Witten invariants for projective surfaces II: semistable case,''
  arXiv:1702.08488 [math.AG].

\bibitem{Witten:1988xi}
  E.~Witten,
  ``Topological Gravity,''
  Phys.\ Lett.\ B {\bf 206} (1988) 601.

\bibitem{Bae:2015eoa}
  J.~Bae, C.~Imbimbo, S.~J.~Rey and D.~Rosa,
  ``New Supersymmetric Localizations from Topological Gravity,''
  JHEP {\bf 1603} (2016) 169
  [arXiv:1510.00006 [hep-th]].

\bibitem{Imbimbo:2018duh}
  C.~Imbimbo and D.~Rosa,
  ``The topological structure of supergravity: an application to supersymmetric localization,''
  arXiv:1801.04940 [hep-th].

\bibitem{Rosa:2018xpt}
  D.~Rosa,
  ``The cohomological structure of generalized Killing spinor equations,''
  arXiv:1801.09347 [hep-th].

\bibitem{Val2018}
	B.~de Wit, S.~Murthy and V.~Reys,
	``BRST quantization and equivariant cohomology: localization with asymptotic boundaries,'' arXiv:1806.03690 [hep-th].

\end{thebibliography}
\end{document}